\documentclass[aps,prd,preprintnumbers,groupedaddress,nofootinbib,amssymb,eqsecnum,longbibliography,10pt]{revtex4-2}

\usepackage{graphicx}
\usepackage{color} 
\usepackage{dcolumn}
\usepackage{amsmath,amsthm,amssymb}
\usepackage{amsfonts}
\usepackage{bm}
\usepackage{mathrsfs}
\usepackage{revsymb}
\usepackage{comment}
\usepackage{braket}
\usepackage{simplewick} 

\usepackage[bookmarks=true,bookmarksnumbered=true,colorlinks=false]{hyperref}
\newif\ifoneauthor
\oneauthortrue

\begin{document}

\title{Emergence of Non-Markovian Classical-Quantum Dynamics from Decoherence}

\author{Shogo Tomizuka}
\email[]{tomizuka@tap.scphys.kyoto-u.ac.jp}
\affiliation{Department of Physics, Kyoto University, Kyoto 606-8502, Japan}

\author{Hiroki Takeda}
\email[]{takeda@tap.scphys.kyoto-u.ac.jp}
\affiliation{The Hakubi Center for Advanced Research, Kyoto University, Kyoto 606-8501, Japan}
\affiliation{Department of Physics, Kyoto University, Kyoto 606-8502, Japan}

\date{\today}

\begin{abstract}
\noindent
The quantum nature of gravity remains experimentally unverified, despite recent proposals to probe it using tabletop experiments such as gravity-mediated entanglement schemes.
In parallel, consistent formulations of classical--quantum dynamics have been developed as alternative descriptions of gravity, in which quantum matter interacts with a classical mediator assumed to be fundamentally classical.
In this work, we show that classical--quantum dynamics arise generically as an effective description of fully quantum systems under decoherence, providing a bridge between fully quantum and classical--quantum dynamics.
We derive the reduced dynamics, which are generically non-Markovian, using an explicit hidden model in which the mediator is coupled to unobserved environmental degrees of freedom.
We identify a concrete criterion for when a classical--quantum interpretation is valid: the semi-Wigner operator associated with the mediator sector must remain positive semidefinite, which can be expressed as a positivity condition on nonlocal kernels governing the evolution.
In the short-memory limit, the reduced evolution reproduces Markovian classical--quantum dynamics of Oppenheim and collaborators.
Our results imply that a classical mediator can arise effectively from decohered quantum dynamics, so that experimental agreement with classical-quantum models does not uniquely determine whether the mediator is fundamentally classical.
\end{abstract}
\maketitle

\section{Introduction}
A consistent unification of quantum theory and gravity is essential for understanding the physics of the early universe and other high-energy regimes. 
However, despite extensive theoretical efforts, there remains no experimental evidence that gravity exhibits quantum behavior, even at low energies.

To address this gap, recent proposals have suggested tabletop experiments to probe the quantum nature of gravity, with the Bose–Marletto–Vedral (BMV) scheme~\cite{Bose:2017nin,Marletto:2017kzi} providing the most prominent example, along with several variants~\cite{Lami:2023gmz}. 
In the BMV setup, two masses are prepared in spatial superpositions and interact gravitationally, leading to entanglement through the accumulation of relative phases.
The observation of such entanglement is often interpreted as evidence that gravity itself must be quantum.
However, it remains unclear which properties of the gravitational field are actually probed and in what precise sense entanglement establishes its quantumness~\cite{Carney:2021vvt,Martin-Martinez:2022uio,Kent:2021ljj,Aziz:2025ypo,Ludescher:2025kko,Galley:2020qsf,Danielson:2021egj,Aziz:2025ypo,Christodoulou:2022mkf}.

In parallel, several models of classical gravity have been developed~\cite{Diosi:1986nu,Penrose:1996cv,Oppenheim:2018igd,Oppenheim:2022xjc,Oppenheim:2023mox,Layton:2022sku,Kafri:2014zsa,DiBartolomeo:2021jpc} as alternative hypotheses against which the predictions of perturbative quantum gravity can be tested.
In these frameworks, the test masses are treated quantum mechanically while the mediating gravitational field remains classical.
Directly coupling classical and quantum systems leads to inconsistencies such as superluminal signaling~\cite{Eppley:1977emg}, which can be avoided by treating the classical sector stochastically~\cite{Galley:2023byb}, leading to consistent formulations such as the Diósi–Penrose model~\cite{Diosi:1986nu,Penrose:1996cv} and the Oppenheim model~\cite{Oppenheim:2018igd,Oppenheim:2022xjc,Oppenheim:2023mox,Layton:2022sku}. 

Despite these developments, classical--quantum dynamics are typically introduced as a fundamental hybrid structure in which the mediator is taken to be intrinsically classical.
However, a fundamentally quantum mediator interacting with unobserved environmental degrees of freedom may decohere and behave effectively classically at the level of reduced dynamics~\cite{Calzetta:1995ys, Martin:1999ih, Takeda:2025cye}.
This suggests that classical--quantum dynamics can arise as an effective description rather than a fundamental one.

In this work, we show that classical--quantum dynamics arise generically as a reduced description of fully quantum systems under decoherence, characterized by a positivity condition on the underlying nonlocal kernels.
We construct a \emph{hidden model} in which the mediator is coupled to additional unobserved degrees of freedom and derive the resulting reduced dynamics by tracing them out.
The resulting evolution is generically non-Markovian, reflecting environmental memory effects, and provides a bridge between fully quantum dynamics and classical--quantum descriptions.

To characterize when a classical--quantum description is valid, we introduce a semi-Wigner representation for the mediator sector, in which the mediator is described in phase space while the remaining sector remains operator-valued.
We show that the reduced dynamics admit a consistent classical--quantum interpretation when the semi-Wigner operator is positive semidefinite, and we identify a positivity condition on the nonlocal kernels governing the evolution.
This provides a concrete and operational criterion for the emergence and consistency of classical--quantum dynamics, and thereby offers a systematic route for deriving such dynamics from microscopic quantum models.

We further analyze the short-memory regime of the non-Markovian dynamics. 
In this limit, we derive a time-local evolution that reproduces a subclass of the Markovian classical--quantum hybrid models proposed by Oppenheim and collaborators~\cite{Oppenheim:2023mox}.
In particular, the classical drift and diffusion, the quantum GKSL structure, and the hybrid couplings arise from local moments of the underlying nonlocal kernels.

These results have implications for interpreting experimental tests of gravitational quantumness. 
Even if an experiment is found to be consistent with classical--quantum dynamics for reduced observables, this does not uniquely determine whether the mediator is fundamentally classical or instead effectively classical through decoherence.

The paper is organized as follows.
Section~\ref{sec:bridge} constructs the hidden model and derives the non-Markovian reduced dynamics in the semi-Wigner representation, while Section~\ref{sec:markov} studies the Markovian case and establishes the connection to the classical--quantum model. 
Section~\ref{sec:discussion} summarizes the results.
Appendix~\ref{sec:example} presents an explicit example of the hidden model, Appendix~\ref{app:trotter} provides the detailed derivation of the Markovian effective action, and Appendix~\ref{sec:CQreview} briefly reviews the classical--quantum framework of Oppenheim.

\section{Construction of Non-Markovian Hidden Model}\label{sec:bridge}
In this section, we construct a hidden model using interacting scalar fields, avoiding technical complications associated with gauge symmetry.
We consider three quantum scalar fields, treating one as an unobserved environment.
Tracing out this environmental field via the influence functional method~\cite{Calzetta:2008iqa,Feynman:1963fq} yields reduced dynamics for the remaining fields, including decoherence effects.
We then introduce a Wigner representation for one of the fields and formulate a criterion under which the resulting reduced description admits a classical--quantum interpretation.

\subsection{General Framework Treating Decoherence}\label{sec:Effac}
We consider three quantum scalar fields, $\psi$, $\phi$, and $h$, with the total action
\begin{align}
    S_\mathrm{tot}&=S_0[\psi]+S_0[h]+S_0[\phi]+S_\mathrm{int}[\psi,h]+S_\mathrm{int}[\psi,\phi]+S_\mathrm{int}[h,\phi]~,\label{eq:Stot}
\end{align}
where the first three terms 
are the free actions and the remaining terms describe the interactions of the form
\begin{align}
    S_\mathrm{int}[\psi,h]&=\int d^4 x ~\lambda_1 F_2[\psi]F_3[h]~,\\
    S_\mathrm{int}[\psi,\phi]&=\int d^4x ~\lambda_2F_2[\psi]G_2[\phi]~,\\
    S_\mathrm{int}[h,\phi]&=\int d^4x~\lambda_3 F_3[h]G_3[\phi]~.
\end{align}
Here $F_2[\psi]$ is a functional of $\psi$ and $G_i[\phi]$ are functionals of $\phi$.
We assume that $F_3[h]$ is an invertible functional of $h$ and introduce a new field variable
\begin{align}
    \tilde h(x)\equiv F_3[h](x)\,,
\end{align}
so that $h$ can be uniquely expressed as $h=F_3^{-1}[\tilde h]$.
This invertibility is necessary because $\tilde h$ will serve as the effective field variable appearing directly in the interaction terms, while the original dynamics must be rewritten in terms of $\tilde h$ through the inverse map.
In terms of $\tilde h$, the interactions become
\begin{align}
    S_\mathrm{int}[\psi,\tilde h]
    =\int d^4x~\lambda_1\,F_2[\psi]\,\tilde h~,\quad S_\mathrm{int}[\tilde h,\phi]
    =\int d^4x~\lambda_3\,\tilde h\,G_3[\phi]~,
\end{align}
while $S_0[h]$ is understood as $S_0[h[\tilde h]]$ with $h[\tilde h]\equiv F_3^{-1}[\tilde h]$.

The full density operator is expressed as
\begin{align}
     \rho[\phi^+,\tilde h^+,\psi^+,\phi^-,\tilde h^-,\psi^-;t]\equiv {}_H\bra{\phi^+,\tilde h^+,\psi^+;t}\hat{\rho}^H\ket{\phi^-,\tilde h^-,\psi^-;t}_H~,
\end{align}
with $\ket{\phi,\tilde h,\psi;t}_H$ denoting simultaneous eigenstates of the Heisenberg picture field operators $\hat{\phi}$, $\widehat{\tilde h}\equiv F_3[\hat h]$, and $\hat{\psi}$.
We treat $\phi$ as an unobserved environment and trace over its degrees of freedom to obtain the reduced density matrix for $\psi$ and $\tilde h$:
\begin{align}
    \rho[\psi^+,\psi^-,\tilde h^c,\tilde h^\Delta;t]\equiv \int d\phi~\rho[\phi,\tilde h^+,\psi^+,\phi,\tilde h^-,\psi^-;t]~,\label{eq:reduced_dens}
\end{align}
where we have introduced the Keldysh variables
\begin{align}
    \tilde h_\Delta\equiv \tilde h_+-\tilde h_-~,\quad\tilde h^c\equiv \frac{1}{2}(\tilde h_++\tilde h_-)~,
\end{align}
so that $\tilde h^\pm=\tilde h^c\pm \tilde h^\Delta/2$.
In this representation, $\tilde h^c$ corresponds to the classical component of the field, capturing the mean trajectory shared by the forward and backward branches, while $\tilde h_\Delta$ describes the quantum fluctuation or coherence between them. 
In the classical–quantum regime discussed below, $\tilde h^c$ plays the role of the classical degree of freedom interacting with the quantum field $\psi$.

Assuming no initial correlations between the three fields, the full density operator factorizes as
\begin{align}
    \hat{\rho}^H=\hat{\rho}^H_\phi\otimes \hat{\rho}^H_{\tilde h}\otimes\hat{\rho}^H_{\psi}~.
\end{align}
The reduced density matrix \eqref{eq:reduced_dens} evolves from $t_i$ to $t_f$ as
\begin{align}
    \rho[\psi^+_f,&\psi_f^-,\tilde h^c_f,\tilde h_f^\Delta;t_f]=\int  d\tilde h_i^cd\tilde h_i^\Delta d\psi_i^+d\psi_i^-\mathcal{J}[\psi^+_f,\psi_f^-,\tilde h^c_f,\tilde h_f^\Delta;t_f|\psi_i^+,\psi_i^-,\tilde h_i^c,\tilde h_i^\Delta;t_i]\rho[\psi_i^+,\psi_i^-,\tilde h_i^c,\tilde h_i^\Delta;t_i]~,\label{eq:evo}
\end{align}
with the propagator given by
\begin{align}
    \mathcal{J}[\psi^+_f,\psi_f^-,\tilde h^c_f,\tilde h_f^\Delta;&t_f|\psi_i^+,\psi_i^-,\tilde h_i^c,\tilde h_i^\Delta;t_i]\nonumber\\
    &=\int^{\psi^+_f}_{\psi_i^+}\mathcal{D}\psi^+\int^{\psi_f^-}_{\psi_i^-}\mathcal{D}\psi^-\int_{\tilde h_i^c}^{\tilde h_f^c}\mathcal{D}\tilde h^c\int_{\tilde h^\Delta_i}^{\tilde h^\Delta_f}\mathcal{D}\tilde h^\Delta \mathcal{J}_{F_3}[\tilde h^+,\tilde h^-]\exp\left[\frac{i}{\hbar}S_\mathrm{eff}[\psi^+,\psi^-,\tilde h^c,\tilde h^\Delta]\right]~.\label{eq:propagator_J}
\end{align}
The Jacobian $\mathcal{J}_{F_3}$ arises from the change of variables $\mathcal{D}h^+\,\mathcal{D}h^- \to \mathcal{D}\tilde h^+\,\mathcal{D}\tilde h^-$ and is formally
\begin{align}
    \mathcal{J}_{F_3}[\tilde h^+,\tilde h^-]
    \equiv \left|\det\!\left(\frac{\delta h^+}{\delta \tilde h^+}\right)\right|
            \left|\det\!\left(\frac{\delta h^-}{\delta \tilde h^-}\right)\right|
    = \left|\det\!\left(\frac{\delta F_3^{-1}[\tilde h^+]}{\delta \tilde h^+}\right)\right|
      \left|\det\!\left(\frac{\delta F_3^{-1}[\tilde h^-]}{\delta \tilde h^-}\right)\right|~.\label{eq:Jacobian}
\end{align}

For notational simplicity, we define
\begin{align}
    &S_0[\tilde h^c,\tilde h^\Delta]\equiv S_0[h[\tilde h^+]]-S_0[h[\tilde h^-]]~,\\
    &S_\mathrm{int}[\psi^+,\psi^-,\tilde h^c,\tilde h^\Delta]\equiv S_\mathrm{int}[\psi^+,h[\tilde h^+]]-S_\mathrm{int}[\psi^-,h[\tilde h^-]]~,
\end{align}
so that the effective action becomes
\begin{align}
    S_\mathrm{eff}[\psi^+,\psi^-,\tilde h^c,\tilde h^\Delta]
    =S_0[\psi^+]-S_0[\psi^-]+S_0[\tilde h^c,\tilde h^\Delta]
    +S_\mathrm{int}[\psi^+,\psi^-,\tilde h^c,\tilde h^\Delta]
    +S_\mathrm{IF}[\psi ^+,\psi^-,\tilde h^c,\tilde h^\Delta]~.\label{eq:Seff}
\end{align}
The influence action $S_\mathrm{IF}$ is defined via the influence functional obtained by tracing out $\phi$,
\begin{align}
    \exp\!\left[\frac{i}{\hbar}S_\mathrm{IF}[\psi^+,\psi^-,\tilde h^c,\tilde h^\Delta]\right]\equiv&\int d\phi_fd\phi_i^+d\phi_i^-\int_{\phi_i^+}^{\phi_f}\mathcal D\phi^+\int_{\phi_i^-}^{\phi_f}\mathcal D\phi^-
    ~\rho_\phi[\phi_i^+,\phi_i^-;t_i]\,\nonumber\\
    &\times\exp\!\left\{\frac{i}{\hbar}\Big(S_0[\phi^+]-S_0[\phi^-]+S_\mathrm{int}[\psi^+,\phi^+]-S_\mathrm{int}[\psi^-,\phi^-]+S_\mathrm{int}[\tilde h^+,\phi^+]-S_\mathrm{int}[\tilde h^-,\phi^-]\Big)\right\},\label{eq:def_SIF}
\end{align}
where $\rho_\phi[\phi_i^+,\phi_i^-;t_i]\equiv {}_H\!\bra{\phi_i^+}\hat\rho_\phi^H\ket{\phi_i^-}_H$ encodes the initial state of the environment field.
When $S_\mathrm{IF}$ is expanded perturbatively in the couplings, linear (tadpole) terms generally appear at first order, e.g.\ terms proportional to $\langle G_2[\phi]\rangle$ or $\langle G_3[\phi]\rangle$.
Such local contributions including possible UV divergences are removed by appropriate counterterms and/or field shifts, and we assume that this renormalization process has been carried out.

We define the noise and dissipation kernels associated with the environment operators $\hat G_I(x)$ ($I=2,3$) by
\begin{align}
    \mathcal{N}_{IJ}(x,y)\equiv 2\lambda_I\lambda_J\big\langle \{\hat G_I(x),\hat G_J(y)\}\big\rangle_\phi,\qquad
    \mathcal{D}_{IJ}(x,y)\equiv -2i\lambda_I\lambda_J\,\theta(x^0-y^0)\big\langle [\hat G_I(x),\hat G_J(y)]\big\rangle_\phi~,
\end{align}
so that $\mathcal{N}_{IJ}(x,y)=\mathcal{N}_{JI}(y,x)$ and $\mathcal{D}_{IJ}(x,y)$ is retarded.

Expanding \eqref{eq:def_SIF} to quadratic order in the couplings, one obtains
\begin{align}
    \frac{i}{\hbar}S_\mathrm{IF}= -\frac{1}{2\hbar^2}\int d^4x d^4y \bigg[&iF_{2\Delta}(x)\mathcal{D}_{22}(x,y)F_{2c}(y)+F_{2\Delta}(x)\mathcal{N}_{22}(x,y)F_{2\Delta}(y)+iF_{2\Delta}(x)\mathcal{D}_{23}(x,y)\tilde h_{c}(y)\nonumber\\
    &+F_{2\Delta}(x)\mathcal{N}_{23}(x,y)\tilde h_{\Delta}(y)+i\tilde h_{\Delta}(x)\mathcal{D}_{32}(x,y)F_{2c}(y)+\tilde h_{\Delta}(x)\mathcal{N}_{32}(x,y)F_{2\Delta}(y)\nonumber\\
    &+i\tilde h_{\Delta}(x)\mathcal{D}_{33}(x,y)\tilde h_{c}(y)+\tilde h_{\Delta}(x)\mathcal{N}_{33}(x,y)\tilde h_{\Delta}(y)\bigg]~.\label{eq:SIF}
\end{align}
Here, $F_{2c}$ and $F_{2\Delta}$ denote the Keldysh components of $F_2[\psi]$,
\begin{align}
    F_{2c}\equiv \frac{1}{2}\big(F_2[\psi^+]+F_2[\psi^-]\big)\,,\qquad F_{2\Delta}\equiv F_2[\psi^+]-F_2[\psi^-]\,.
    \label{eq:Keldysh_F_2}
\end{align} 

In Eq.~\eqref{eq:SIF}, the noise kernels $\mathcal{N}_{22}$ and $\mathcal{N}_{33}$ govern the decoherence of $\psi$ and $\tilde h$, respectively, while the dissipation kernels $\mathcal{D}_{22}$ and $\mathcal{D}_{33}$ encode their dissipative dynamics.
The cross-kernels $\mathcal{N}_{23},\mathcal{D}_{23}$ (and $\mathcal{N}_{32},\mathcal{D}_{32}$) describe correlated noise and backreaction between the two sectors induced by the common environment.
When the environment is close to equilibrium, these structures are often related by fluctuation-dissipation relations.
If $\mathcal{N}_{33}$ is sufficiently large compared to the relevant dynamical scales, quantum coherences in $\tilde h$ are strongly suppressed and the field can behave effectively as a classical field. 
 
\subsection{Wigner transformation for one field}

The reduced dynamics described above interpolates between fully quantum and classical–quantum regimes.
To determine where the system lies along this continuum, it is useful to perform a Wigner transformation with respect to the degree of freedom whose classicality is to be assessed.
In the present construction, we apply this transformation to the field variable $\tilde h\equiv F_3[h]$.
The resulting criterion therefore characterizes effective classicality in the $\tilde h$ sector.
\footnote{This notion of classicality is tied to the phase-space variables $(\tilde h,\tilde \pi)$ associated with the interaction variable $\tilde h=F_3[h]$. In general, nonnegativity of the Wigner function in these variables does not imply nonnegativity of the Wigner function in the original variables $(h,\pi)$, even when $F_3$ is invertible, because the Wigner transform is covariant only under linear canonical transformations, not under a generic nonlinear field redefinition. Thus, for nonlinear $F_3$, nonnegativity of the Wigner function should be interpreted as effective classicality of the sector coupled through $F_3[h]$, rather than of the original field variable $h$ itself.}

We therefore introduce the Wigner transform of the reduced density matrix with respect to $\tilde h$ as
\begin{align}
    \bra{\psi^+}\hat{W}(\tilde h^c,\tilde \pi^c)\ket{\psi^-}
    &\equiv \int d\tilde h^\Delta 
    \exp\!\left(-\frac{i}{\hbar}\tilde h^\Delta\tilde \pi^c\right)
    \rho[\psi^+,\psi^-,\tilde h^c,\tilde h^\Delta]~,
\end{align}
where $\tilde \pi$ is the momentum conjugate to $\tilde h$.  
The resulting operator $\hat{W}(\tilde h^c, \tilde \pi^c)$ acts on the Hilbert space of $\psi$, and we therefore refer to it as the \emph{semi-Wigner operator}.
Tracing over $\psi$ yields the phase-space distribution associated with $\tilde h$,
\begin{align}
    p(\tilde h^c,\tilde \pi^c)
    =\int d\psi\,\bra{\psi}\hat{W}(\tilde h^c,\tilde \pi^c)\ket{\psi}~.
\end{align}
This function can in general take negative values, signaling the nonclassical nature of $\tilde h$.

Finally, applying the same Wigner transformation to the evolution equation~\eqref{eq:evo} yields the path-integral representation of the semi-Wigner operator:
\begin{align}
    \bra{\psi_f^+}\hat{W}(\tilde h_f^c,\tilde \pi^c_f)\ket{\psi_f^-}
    &=\int d\tilde h_f^\Delta 
    \exp\bigg(-\frac{i}{\hbar}\tilde h_f^\Delta\tilde \pi_f^c\bigg)
    \rho[\psi_f^+,\psi_f^-,\tilde h_f^c,\tilde h_f^\Delta;t_f]\nonumber\\
    &=\int d\tilde h_i^c\,d\tilde \pi_i^c\,d\psi_i^+\,d\psi_i^-\, 
    \mathcal{J}_\mathrm{eff}^\mathrm{W}
    [\psi_f^+,\psi_f^-,\tilde h_f^c,\tilde \pi_f^c;t_f
    |\psi_i^+,\psi_i^-,\tilde h_i^c,\tilde \pi_i^c;t_i]\,
    \bra{\psi_i^+}\hat{W}(\tilde h_i^c,\tilde \pi_i^c)\ket{\psi_i^-}~,\label{eq:path_int_semi_wigner}
\end{align}
where the kernel $\mathcal{J}_\mathrm{eff}^\mathrm{W}$ is given by
\begin{align}
    \mathcal{J}_\mathrm{eff}^\mathrm{W}[\psi_f^+,\psi_f^-,\tilde h_f^c,\tilde\pi_f^c;t_f|\psi_i^+,\psi_i^-,\tilde h_i^c,\tilde\pi_i^c;t_i]=\int_{\psi_i^+}^{\psi_f^+}\mathcal{D}\psi^+\int_{\psi_i^-}^{\psi_f^-}\mathcal{D}\psi^-\int_{\tilde h_i^c}^{\tilde h_f^c}\mathcal{D}\tilde h^c\,\exp\!\left[\frac{i}{\hbar}S_\mathrm{eff}^\mathrm{W}[\psi^+,\psi^-,\tilde h^c;\tilde\pi_i^c,\tilde\pi_f^c]\right]~.\label{eq:JWeff_def}
\end{align}
Here the Wigner-transformed effective action $S_\mathrm{eff}^\mathrm{W}$ is defined by
\begin{align}
    \exp\!\left[\frac{i}{\hbar}S_\mathrm{eff}^\mathrm{W}[\psi^+,\psi^-,\tilde h^c;\tilde\pi_i^c,\tilde\pi_f^c]\right]\equiv\int d\tilde h_f^\Delta\, d\tilde h_i^\Delta\,\exp\!\left(-\frac{i}{\hbar}\tilde h_f^\Delta\tilde\pi_f^c+\frac{i}{\hbar}\tilde h_i^\Delta\tilde\pi_i^c\right)\int_{\tilde h_i^\Delta}^{\tilde h_f^\Delta}\mathcal{D}\tilde h^\Delta\,\exp\!\left[\frac{i}{\hbar}S_\mathrm{eff}[\psi^+,\psi^-,\tilde h^c,\tilde h^\Delta]\right]~.\label{eq:SeffW_def}
\end{align}

Using the path-integral representation of $\mathcal{J}$ in Eq.~\eqref{eq:propagator_J} and the quadratic influence action in Eq.~\eqref{eq:SIF}, the $\tilde h^\Delta$-dependence of the exponent takes the Gaussian form
\begin{align}
    \frac{i}{\hbar}S_\mathrm{eff}[\psi^+,\psi^-,\tilde h^c,\tilde h^\Delta]
    \supset\frac{i}{\hbar}\int d^4x~\tilde h^\Delta(x)\,\mathcal{E}(x)-\frac{1}{2\hbar^2}\int d^4x\,d^4y~\tilde h^\Delta(x)\,\mathcal{N}_{33}^R(x,y)\,\tilde h^\Delta(y)~,\label{eq:hDelta_gauss}
\end{align}
where
\begin{align}
    \mathcal{E}(x)
    &\equiv\left.\frac{\delta}{\delta \tilde h(x)}S_0[h[\tilde h]]\right|_{\tilde h=\tilde h^c}+\lambda_1 F_{2c}(x)-\frac{1}{2\hbar}\int d^4y\Big[\mathcal{D}_{33}(x,y)\tilde h^c(y)+\mathcal{D}_{32}(x,y)F_{2c}(y)\Big]-\frac{i}{\hbar}\int d^4y~\mathcal{N}_{32}(x,y)F_{2\Delta}(y)~,
    \label{eq:E_def}
\end{align}
and $\mathcal{N}_{33}^R$ denotes the renormalized noise kernel.\footnote{
We expand the Jacobian factor $\mathcal J_{F_3}[\tilde h^+,\tilde h^-]$ in Eq.~\eqref{eq:Jacobian} in powers of $\tilde h_\Delta\equiv \tilde h^+-\tilde h^-$.
Using $\tilde h^\pm=\tilde h_c\pm \tfrac12\tilde h_\Delta$ and denoting
$\Phi[\tilde h]\equiv \ln\!\left|\det\!\left(\delta F_3^{-1}[\tilde h]/\delta\tilde h\right)\right|$,
we have
\begin{align}
\ln \mathcal J_{F_3}[\tilde h^+,\tilde h^-]=\Phi[\tilde h^+]+\Phi[\tilde h^-]
=2\Phi[\tilde h_c]+\frac14\!\int\! dxd y\;\Phi^{(2)}_{xy}[\tilde h_c]\,\tilde h_\Delta(x)\tilde h_\Delta(y)+\mathcal O(\tilde h_\Delta^4),\qquad\Phi^{(2)}_{xy}\equiv \frac{\delta^2\Phi}{\delta \tilde h(x)\delta \tilde h(y)}.\label{eq:jac_taylor}
\end{align}
In the strong-decoherence regime, the path integral is localized near $\tilde h_\Delta=0$ by the noise term
$\exp\{-\tfrac12\int \tilde h_\Delta\,\mathcal N_{33}\,\tilde h_\Delta\}$, so that the expansion in Eq.~\eqref{eq:jac_taylor} is well controlled and the higher-order terms $\mathcal O(\tilde h_\Delta^4)$ may be neglected.
The quadratic Jacobian contribution can therefore be incorporated into the full quadratic kernel in the $\tilde h_\Delta$ sector as
\begin{align}
\ln \mathcal J_{F_3}[\tilde h^+,\tilde h^-]=2\Phi[\tilde h_c]-\frac{1}{2\hbar^2}\!\int\! dxd y\;\tilde h_\Delta(x)\,\mathcal N^{J}_{33}(x,y)\,\tilde h_\Delta(y)+\mathcal O(\tilde h_\Delta^4),\qquad\mathcal N^{J}_{33}(x,y)\equiv -\frac{\hbar^2}{2}\,\Phi^{(2)}_{xy}[\tilde h_c].\label{eq:jac_noise}
\end{align}
In what follows, we restrict attention to invertible point transformations,
\[
\tilde h(x)=F_3(h(x)),
\qquad
h(x)=F_3^{-1}(\tilde h(x)),
\]
for which the transformation at each spacetime point depends only on the field value at the same point and involves no derivatives.
For this class of transformations, the second functional derivative is local,
\[
\Phi^{(2)}_{xy}[\tilde h_c]=\phi_2(\tilde h_c;x)\,\delta(x-y),
\]
for some local coefficient $\phi_2(\tilde h_c;x)$.
In general, the coefficient $\phi_2(\tilde h_c;x)$ contains a UV-divergent contact term.
To absorb this divergent local contribution, we introduce a local counterterm kernel $\mathcal N^{\rm ct}_{33}(x,y)$ with the same local structure.
The finite renormalized kernel is then defined by
\begin{align}
\mathcal N_{33}^R(x,y)\equiv\mathcal N_{33}(x,y)+\mathcal N^{J}_{33}(x,y)+\mathcal N^{\rm ct}_{33}(x,y),\label{eq:N33R}
\end{align}
so that the divergent local part of the Jacobian contribution is absorbed into the counterterm and $\mathcal N_{33}^R$ is finite.
}

Introducing an auxiliary noise field $\xi$ through a Hubbard--Stratonovich transformation,
\begin{align}
    \exp\!\left[-\frac{1}{2\hbar^2}\tilde h^\Delta\cdot \mathcal{N}_{33}^R\cdot \tilde h^\Delta\right]\propto\int \mathcal{D}\xi~\exp\!\left[-\frac{1}{2}\xi\cdot (\mathcal{N}_{33}^R)^{-1}\cdot \xi+\frac{i}{\hbar}\tilde h^\Delta\cdot \xi\right]~,\label{eq:HS}
\end{align}
where the dot denotes spacetime convolution, and $(\mathcal N_{33}^R)^{-1}$ denotes the inverse kernel defined by
\begin{align}
    \int d^4z\,\mathcal N_{33}^R(x,z)\,(\mathcal N_{33}^R)^{-1}(z,y)=\delta^{(4)}(x-y)~,
\end{align}
on the support of $\mathcal N_{33}^R$.
The functional integral over $\tilde h^\Delta$ then becomes linear and can be carried out explicitly.
Up to an overall normalization, one obtains
\begin{align}
    \exp\!\left[\frac{i}{\hbar}S_\mathrm{eff}^\mathrm{W}[\psi^+,\psi^-,\tilde h^c;\tilde\pi_i^c,\tilde\pi_f^c]\right] 
    &\propto\int \mathcal{D}\xi~\exp\!\left[-\frac{1}{2}\xi\cdot (\mathcal{N}_{33}^R)^{-1}\cdot \xi\right]\,\delta\!\Big[\mathcal{E}+\xi\Big]\,\delta\!\Big[\tilde\Pi_f^c-\tilde\pi_f^c\Big]\,\delta\!\Big[\tilde\Pi_i^c-\tilde\pi_i^c\Big]\,\exp\!\left[\frac{i}{\hbar}S_{\rm rest}[\psi^+,\psi^-,\tilde h^c]\right]~,\label{eq:SeffW_from_delta}
\end{align}
where the bulk delta functional $\delta[\mathcal{E}+\xi]$ arises because, after the Hubbard--Stratonovich transformation, the exponent is linear in $\tilde h^\Delta$ in the interior of the time contour, so that the functional integration over $\tilde h^\Delta$ imposes the constraint $\mathcal{E}+\xi=0$.
The endpoint delta functionals $\delta[\tilde\Pi_f^c-\tilde\pi_f^c]$ and $\delta[\tilde\Pi_i^c-\tilde\pi_i^c]$ originate from the boundary terms in the variation of the action with respect to $\tilde h^\Delta$, together with the endpoint Fourier factors appearing in Eq.~\eqref{eq:SeffW_def}.
Here
\begin{align}
    \tilde\Pi(x)\equiv\frac{\partial \mathcal{L}_0[h[\tilde h]]}{\partial \dot{\tilde h}(x)}~,
\end{align}
is the canonical momentum conjugate to the redefined field $\tilde h$, where $\mathcal{L}_0$ is the free Lagrangian density defined by $S_0[h]=\int d^4x\,\mathcal{L}_0[h]$.

The functional $S_{\rm rest}$ collects all terms that are independent of $\tilde h^\Delta$:
\begin{align}
    \frac{i}{\hbar}S_{\rm rest}&=\frac{i}{\hbar}\big(S_0[\psi_+]-S_0[\psi_-]\big)+\frac{i}{\hbar}\int d^4x ~\lambda_1 F_{2\Delta}\tilde h_c\nonumber\\
    &\hspace{5mm}-\frac{1}{2\hbar^2}\int d^4x d^4y \bigg[iF_{2\Delta}(x)\mathcal{D}_{22}(x,y)F_{2c}(y)+F_{2\Delta}(x)\mathcal{N}_{22}(x,y)F_{2\Delta}(y)+iF_{2\Delta}(x)\mathcal{D}_{23}(x,y)\tilde h_{c}(y)\bigg]~.
\end{align}
Equivalently, integrating out $\xi$ yields the configuration-space form
\begin{align}
    \exp\!\left[\frac{i}{\hbar}S_\mathrm{eff}^\mathrm{W}[\psi^+,\psi^-,\tilde h^c;\tilde\pi_i^c,\tilde\pi_f^c]\right]
    &\propto\delta\!\Big[\tilde\Pi_f^c-\tilde\pi_f^c\Big]\,\delta\!\Big[\tilde\Pi_i^c-\tilde\pi_i^c\Big]\,\exp\!\left[-\frac{1}{2}\mathcal{E}\cdot (\mathcal{N}_{33}^R)^{-1}\cdot \mathcal{E}\right]\,\exp\!\left[\frac{i}{\hbar}S_{\rm rest}[\psi^+,\psi^-,\tilde h^c]\right]~.\label{eq:SeffW_gauss}
\end{align}
Substituting Eq.~\eqref{eq:SeffW_gauss} into Eq.~\eqref{eq:JWeff_def}, and then inserting the result into Eq.~\eqref{eq:path_int_semi_wigner}, we obtain the time evolution of the semi-Wigner operator $\hat W(\tilde h^c,\tilde \pi^c)$.
In this way, the reduced dynamics are described as an operator evolution in which the $\tilde h$ sector is represented in phase space, while the $\psi$ sector is kept in Hilbert-space operator form.
Thus, for a given initial semi-Wigner operator at $t_i$, the effective action $S_\mathrm{eff}^\mathrm{W}$ determines its propagation to $t_f$ through the path integral.
This provides the starting point for analyzing when the reduced dynamics admit an effective classical-quantum interpretation.

\subsection{The Condition for Classical–Quantum Description}
We now consider the regime in which the phase-space distribution satisfies
\begin{align}
    p(\tilde h^c,\tilde \pi^c)\ge0~,
\end{align}
for all $(\tilde h^c,\tilde \pi^c)$.
In this regime, $p(\tilde h^c,\tilde \pi^c)$ can be interpreted as a genuine classical probability distribution on phase space.  
A sufficient condition for this is that the semi-Wigner operator be positive semidefinite on the Hilbert space of $\psi$,
\begin{align}
    \bra{\psi}\hat{W}(\tilde h^c,\tilde \pi^c)\ket{\psi}\ge0  ~,\label{eq:pos.cond.}
\end{align}
for every state $\ket{\psi}$.
For each fixed $(\tilde h^c,\tilde \pi^c)$, Eq.~\eqref{eq:pos.cond.} implies that $\hat W(\tilde h^c,\tilde \pi^c)$ is a positive trace-class operator on the Hilbert space of $\psi$ with trace $p(\tilde h^c,\tilde \pi^c)$, so that whenever $p(\tilde h^c,\tilde \pi^c)>0$, the normalized operator $\hat W(\tilde h^c,\tilde \pi^c)/p(\tilde h^c,\tilde \pi^c)$ defines the conditional density operator of $\psi$.
When this condition is satisfied, the subsystem $\tilde h$ behaves effectively as a classical degree of freedom.  
In this sense, a positive semidefinite semi-Wigner operator represents a \emph{classical–quantum state}, even though the underlying theory remains fully quantum.  

However, even if the semi-Wigner operator becomes positive semidefinite at some moment, rendering $\tilde h$ effectively classical at that time, this property is not generically preserved under subsequent evolution. 
Because of its coupling to $\psi$, the operator $\hat{W}$ may later cease to be positive semidefinite, which in its phase-space representation corresponds to the reappearance of negative regions.

To identify regimes in which $\hat W(\tilde h^c,\tilde\pi^c)$ remains positive semidefinite throughout the evolution, it is useful to recast the semi-Wigner propagator into a Kraus-type form.
We start from Eq.~\eqref{eq:SeffW_gauss}, in which the $\tilde h^\Delta$ and $\xi$ integrations have already been carried out.
In the following discussion, we suppress the endpoint delta functionals $\delta[\tilde\Pi_f^c-\tilde\pi_f^c]\,\delta[\tilde\Pi_i^c-\tilde\pi_i^c]$, since they do not affect the branch structure relevant to the decomposition below.
We therefore write
\begin{align}
    \exp\!\left[\frac{i}{\hbar}S_{\rm eff}^{\rm W}[\psi^+,\psi^-,\tilde h^c]\right]
    \propto
    \exp\!\left[-\frac12\,\mathcal E\cdot(\mathcal N_{33}^R)^{-1}\cdot\mathcal E\right]\,
    \exp\!\left[\frac{i}{\hbar}S_{\rm rest}[\psi^+,\psi^-,\tilde h^c]\right].
    \label{eq:SeffW_start_CQ}
\end{align}

Using Eq.~\eqref{eq:E_def}, we decompose the functional $\mathcal E$ that defines the deterministic part of the effective stochastic equation as
\begin{align}
    \mathcal E= \mathcal A[\tilde h^c]+\mathcal L\cdot F_{2c}-\frac{i}{\hbar}\,\mathcal N_{32}\cdot F_{2\Delta},\label{eq:E_decompose_CQ}
\end{align}
where
\begin{align}
    \mathcal A[\tilde h^c](x)
    &\equiv
    \left.\frac{\delta}{\delta \tilde h(x)}S_0[h[\tilde h]]\right|_{\tilde h=\tilde h^c}
    -\frac{1}{2\hbar}\,(\mathcal D_{33}\cdot \tilde h^c)(x),\\
    \mathcal L(x,y)
    &\equiv
    \lambda_1 \delta(x-y)-\frac{1}{2\hbar}\mathcal D_{32}(x,y)~.
\end{align}
Introducing
\begin{align}
    \mathcal Q\equiv (\mathcal N_{33}^R)^{-1},
    \qquad
    \mathcal B_\pm \equiv \frac12\,\mathcal L \mp \frac{i}{\hbar}\,\mathcal N_{32},
\end{align}
we may rewrite $\mathcal E$ as
\begin{align}
    \mathcal E
    =\mathcal A[\tilde h^c]+\mathcal B_+\cdot F_2[\psi_+]+\mathcal B_-\cdot F_2[\psi_-]~.
\end{align}
Substituting this expression into Eq.~\eqref{eq:SeffW_start_CQ}, we can reorganize the effective action according to its dependence on the two Keldysh branches.
More precisely, the effective action naturally separates into three parts:
a purely classical contribution depending only on the trajectory $\tilde h^c$, a pair of single-branch terms that govern the evolution on each branch separately, and a residual term that couples the forward and backward branches.
Accordingly, we write
\begin{align}
    \frac{i}{\hbar}S_{\rm eff}^{\rm W}[\psi^+,\psi^-,\tilde h^c]
    =
    -I_C[\tilde h^c]
    +I_{CQ}[\tilde h^c,\psi_+]
    +I_{CQ}^\ast[\tilde h^c,\psi_-]
    +\tilde I_{CQ}[\tilde h^c,\psi_+,\psi_-]~.
    \label{eq:split_full_CQ}
\end{align}
The first term is the purely classical weight associated with the phase-space trajectory of $\tilde h$,
\begin{align}
    I_C[\tilde h^c]
    \equiv
    \frac12\,\mathcal A[\tilde h^c]\cdot \mathcal Q\cdot \mathcal A[\tilde h^c]~,
    \label{eq:IC_def_section}
\end{align}
while the second term collects all contributions that depend only on a single branch,
\begin{align}
    I_{CQ}[\tilde h^c,\psi]
    &\equiv
    \frac{i}{\hbar}S_0[\psi]
    +\frac{i}{\hbar}\lambda_1\,\tilde h^c\cdot F_2[\psi]
    -\frac{i}{2\hbar^2}\,F_2[\psi]\cdot \mathcal D_{23}\cdot \tilde h^c
    -\mathcal A[\tilde h^c]\cdot \mathcal Q\cdot \mathcal B_+\cdot F_2[\psi]\nonumber\\
    &\qquad
    -\frac12\,F_2[\psi]\cdot
    \left(
        \mathcal B_+^T\cdot \mathcal Q\cdot \mathcal B_+
        +\frac{1}{\hbar^2}\mathcal N_{22}
        +\frac{i}{2\hbar^2}\mathcal D_{22}^{\,s}
    \right)\cdot F_2[\psi]~,
    \label{eq:ICQ_explicit}
\end{align}
where $K^T(x,y)\equiv K(y,x)$ denotes the transpose of a kernel and
\begin{align}
    \mathcal D_{22}^{\,s}(x,y)\equiv \frac12\Big(\mathcal D_{22}(x,y)+\mathcal D_{22}(y,x)\Big)~,
\end{align}
is the symmetric part of the dissipation kernel.
The remaining term $\tilde I_{CQ}$ contains the genuine coupling between the two branches and can be written in the bilinear form
\begin{align}
    \tilde I_{CQ}[\tilde h^c,\psi_+,\psi_-]
    \equiv
    \int d^4x\,d^4y~
    F_2[\psi_+](x)\,C(x,y)\,F_2[\psi_-](y)~,
    \label{eq:Itilde_def_section}
\end{align}
where
\begin{align}
    C
    =
    \frac{1}{\hbar^2}\,\mathcal N_{22}
    -\frac{i}{2\hbar^2}\,\mathcal D_{22}^{\,a}
    -\mathcal B_+^T\cdot \mathcal Q\cdot \mathcal B_-,
    \label{eq:C_compact_section}
\end{align}
and
\begin{align}
    \mathcal D_{22}^{\,a}(x,y)\equiv \frac12\Big(\mathcal D_{22}(x,y)-\mathcal D_{22}(y,x)\Big)~,
\end{align}
is the antisymmetric part of $\mathcal D_{22}$.
Using $\mathcal N_{23}(x,y)=\mathcal N_{32}(y,x)$ together with the symmetry of $\mathcal Q$, one finds that
\begin{align}
    C(x,y)=C^\ast(y,x),
\end{align}
so that $C$ is Hermitian as a kernel.

We now unravel the bilinear coupling \eqref{eq:Itilde_def_section} by introducing a complex auxiliary noise $\eta$,
\begin{align}
    \exp\!\big[\tilde I_{CQ}\big]=\int \mathcal D\eta \mathcal D\eta^{\ast}~P[\eta]\,\exp\!\left[ i\int d^4x~\eta(x)\,F_2[\psi_+](x)-i\int d^4x~\eta^\ast(x)\,F_2[\psi_-](x) \right],\label{eq:unravel_Itilde_section}
\end{align}
with the Gaussian weight
\begin{align}
    P[\eta]\propto \exp\!\left[-\int d^4x\,d^4y~\eta^\ast(x)\,C^{-1}(x,y)\,\eta(y)\right].\label{eq:Peta_section}
\end{align}
Defining the (generally non-unitary) noise-resolved operator on the Hilbert space of $\psi$ by the single-branch path integral
\begin{align}
    \langle\psi^f|\hat K[\eta;\tilde h^c]|\psi^i\rangle\equiv\int_{\psi^i}^{\psi^f}\mathcal D\psi~\exp\!\left[ I_{CQ}[\tilde h^c,\psi] +i\int d^4x~\eta(x)\,F_2[\psi](x)\right],\label{eq:Ueta_section}
\end{align}
the semi-Wigner evolution can be written in the Kraus form. 
Restoring the endpoint constraints implied by the Wigner transform, we obtain
\begin{align}
    &\langle\psi_+^f|\hat W_f(\tilde h_f^c,\tilde\pi_f^c)|\psi_-^f\rangle \nonumber\\
    &=\int\mathcal D\eta\mathcal D\eta^{\ast}~P[\eta]\int d\tilde h_i^c\,d\tilde\pi_i^c\int_{\tilde h_i^c}^{\tilde h_f^c}\mathcal D\tilde h^c~e^{-I_C[\tilde h^c]}\, \delta\!\Big[\tilde\Pi_f^c-\tilde\pi_f^c\Big]\,\delta\!\Big[\tilde\Pi_i^c-\tilde\pi_i^c\Big]\,\langle\psi_+^f|\hat K[\eta;\tilde h^c]\,\hat W_i(\tilde h_i^c,\tilde\pi_i^c)\, \hat K^\dagger[\eta;\tilde h^c]|\psi_-^f\rangle~.\label{eq:randomKraus_section}
\end{align}
If $P[\eta]$ is a genuine probability functional, Eq.~\eqref{eq:randomKraus_section} is a convex mixture of completely positive maps and therefore preserves positivity:
if \eqref{eq:pos.cond.} holds initially for all phase-space points, then it remains valid at all later times.

A sufficient condition for completely positivity is that the kernel $C$ be positive in the sense of bilinear forms,
\begin{align}
    \int d^4x\,d^4y~f^\ast(x)\,C(x,y)\,f(y)\ge0~,\label{eq:CP_C_section}
\end{align}
for all test functions $f$.
In addition, convergence of the classical weight requires positivity of $\mathcal N_{33}^R$ on its support:
\begin{align}
    \int d^4x\,d^4y~g(x)\,\mathcal N_{33}^R(x,y)\,g(y)\ge0~,
\end{align}
for all test functions $g$.
Under these conditions, $\hat W(\tilde h^c,\tilde\pi^c)$ remains positive semidefinite throughout the evolution, and the phase-space distribution stays nonnegative.
In such situations, Eq.~\eqref{eq:SeffW_gauss} defines a consistent effective action for a classical--quantum description.

Unlike existing classical--quantum models formulated in a Markovian, time-local form, the hidden model constructed here is generically non-Markovian, and the relevant positivity requirement is therefore a condition on the kernels $C(x,y)$ and $\mathcal N_{33}^R(x,y)$ over the whole evolution interval.
In particular, complete positivity does not require a positivity condition at each instant.
Once the interaction structure is specified, the explicit form of the kernels $\mathcal N_{IJ},\mathcal D_{IJ}$ determines $C$ via \eqref{eq:C_compact_section}, and the corresponding classical--quantum propagator $\mathcal J_{\rm CQ}$ and its positivity condition can be worked out explicitly in a concrete model, which we present in Appendix~\ref{sec:example}.

\section{Markovian hidden model}\label{sec:markov}
In Sec.~\ref{sec:bridge}, we derived a non-Markovian hidden-model description in which the positivity of the semi-Wigner operator is controlled by the kernel condition \eqref{eq:CP_C_section}. 
We now consider the short-memory regime of this construction and show that the resulting dynamics realizes a restricted class of the Markovian classical-quantum dynamics. 
In particular, the classical drift and diffusion, the quantum GKSL block, and the hybrid couplings all arise from the local moment expansion of the nonlocal kernels. 

The order of steps used in Sec.~\ref{sec:bridge} cannot be applied directly in the Markovian analysis.
Inserting the local expansion \eqref{eq:disc_markovN_local} into the unreduced path integral generates, after integration by parts, a contribution containing $\dot{\tilde h}_\Delta^{\,2}$.
As a result, the bulk integration over $\tilde h_\Delta$ is then no longer the linear constraint-generating integral that led to Eq.~\eqref{eq:SeffW_from_delta}. 
Instead, it produces the inverse of a differential operator of the form $N_{33}-N_{33}^{(2)}\partial_t^2$, thereby obscuring the local structure that the Markov approximation is meant to isolate.
For this reason, in the short-memory regime it is more natural to first derive the local master equation and then reconstruct from it the corresponding local effective action by a Trotter decomposition.

In this section, for notational simplicity, we write $(\tilde h,\tilde\pi)$ instead of $(\tilde h^c,\tilde\pi^c)$.

\subsection{Markov approximation and master equation}
The Markov approximation is implemented by expanding the nonlocal kernels around equal times. 
For the noise kernels, symmetry under interchange of the two arguments implies that the first nontrivial derivative correction is the second time moment. 
For the dissipation kernels, the leading derivative correction is the first time moment. 
We therefore write
\begin{align}
\mathcal N_{IJ}(x,y)&\simeq2N_{IJ}\,\delta^{(4)}(x-y)-2N_{IJ}^{(2)}\,\delta^{(3)}(\mathbf x-\mathbf y)\,\partial_{t_x}^2\delta(t_x-t_y),\label{eq:disc_markovN_local}\\
\mathcal D_{IJ}(x,y)&\simeq2D_{IJ}\,\delta^{(4)}(x-y)+2D_{IJ}^{(1)}\,\delta^{(3)}(\mathbf x-\mathbf y)\,\partial_{t_x}\delta(t_x-t_y).\label{eq:disc_markovD_local_D}
\end{align}
Here $N_{33}$ and $N_{33}^{(2)}$ denote the local moments of the renormalized kernel $\mathcal N_{33}^R$ introduced in~\eqref{eq:hDelta_gauss}. 
For simplicity, we suppress the superscript $R$ on these local coefficients.\footnote{For a stationary thermal environment, the noise and dissipative kernels are not independent. 
The KMS condition implies the fluctuation-dissipation relation in frequency space. 
In the high-temperature and low-frequency limit $\beta\hbar\omega\ll1$, one finds the relation
\begin{align}
N_{IJ}\simeq \frac{4T}{\hbar}\,D_{IJ}^{(1)}.
\label{eq:disc_FDR_local}
\end{align}
Thus thermal equilibrium ties the leading local noise coefficient to the first dissipative moment. 
By contrast, $N_{IJ}^{(2)}$ is not fixed by $D_{IJ}^{(1)}$ alone, since it depends also on higher odd moments of the retarded kernel.
}
It should be noted that $D_{IJ}$ and $D_{IJ}^{(1)}$ correspond to the symmetric and antisymmetric parts of the local expansion of the dissipation kernel, denoted by $\mathcal D_{IJ}^{\,s}$ and $\mathcal D_{IJ}^{\,a}$, respectively.

It is also convenient to introduce
\begin{align}
\hat F_2(\mathbf x)\equiv F_2[\hat\psi](\mathbf x),\qquad\hat R_2(\mathbf x)\equiv \frac{i}{\hbar}[\hat H_\psi,\hat F_2(\mathbf x)],\label{eq:disc_markov_F2R2}
\end{align}
where $\hat R_2$ corresponds to $\partial _tF_2$ in the path integral representation.
The second time moment of $\mathcal N_{22}$ gives rise to terms quadratic in $\hat R_2$, while the first time moment of $\mathcal D_{22}$ mixes $\hat F_2$ and $\hat R_2$. 
For this reason, $(\hat F_2,\hat R_2)$ is the natural local operator basis in the Markovian regime.

We now derive the local master equation.
Because integrating out $\tilde h_\Delta$ introduces the inverse kernel $(N_{33}^R)^{-1}$ and thus leads to a nonlocal reduced structure, the Markovian reduction must be performed at the level of the unreduced effective action~\eqref{eq:Seff}.
The local generator is obtained by inserting \eqref{eq:disc_markovN_local} and \eqref{eq:disc_markovD_local_D} into the exponent~\eqref{eq:Seff}, expanding the short-time propagator to first order, and performing the partial Wigner transform in the $\tilde h$ sector.

Under this transformation, $\tilde h_\Delta$ and $\dot{\tilde h}_\Delta$ are represented by functional derivatives with respect to $\tilde\pi$ and $\tilde h$.
To make contact with the time-local CQ master equation, we restrict attention to models for which the free part $S_0[h[\tilde h]]$ gives rise to a local Markovian generator that is at most quadratic in the phase-space variables $(\tilde h,\tilde\pi)$, so that the Markov approximation closes on a time-local evolution equation with at most drift and diffusion terms in the $\tilde h$ sector.
In the $\psi$ sector, the Keldysh combinations~\eqref{eq:Keldysh_F_2} and their time derivatives are mapped to operator actions on $\hat W$.
Since $F_2[\psi^+]$ and $F_2[\psi^-]$ act on the forward and backward branches, they correspond to left and right multiplication by $\hat F_2$, respectively.
Therefore these combinations are represented by
\begin{align}
F_{2c}\rightarrow \frac12\{\hat F_2,\cdot\},
\qquad
F_{2\Delta}\rightarrow [\hat F_2,\cdot],
\qquad
\partial_tF_{2c}\rightarrow \frac12\{\hat R_2,\cdot\},
\qquad
\partial_tF_{2\Delta}\rightarrow [\hat R_2,\cdot].
\label{eq:app_correspondence_rules}
\end{align}
The master equation for $\hat W$ then takes the form
\begin{align}
\partial_t \hat W[\tilde h,\tilde\pi]&=-\int d^3\mathbf x\,\frac{\delta}{\delta \tilde h(\mathbf x)}\Bigl(\tilde\pi(\mathbf x)\hat W\Bigr)-\int d^3\mathbf x\,\frac{\delta}{\delta\tilde\pi(\mathbf x)}\Bigl(\tilde A[\tilde h,\tilde\pi](\mathbf x)\hat W\Bigr)\nonumber\\
&\quad+\int d^3\mathbf x\,N_{33}^{(2)}\frac{\delta^2\hat W}{\delta\tilde h(\mathbf x)^2}+\int d^3\mathbf x\,N_{33}\frac{\delta^2\hat W}{\delta\tilde\pi(\mathbf x)^2}-\frac{i}{\hbar}\Bigl[\hat H_{\rm eff}[\tilde h,\tilde\pi],\hat W\Bigr]\nonumber\\
&\quad+\mathcal L_{22}^{\rm GKSL}[\hat W]+\int d^3\mathbf x\,\gamma\,\Bigl\{\hat F_2(\mathbf x),\frac{\delta\hat W}{\delta\tilde\pi(\mathbf x)}\Bigr\}-\int d^3\mathbf x\,i\nu\,\Bigl[\hat F_2(\mathbf x),\frac{\delta\hat W}{\delta\tilde\pi(\mathbf x)}\Bigr]\nonumber\\
&\quad+\int d^3\mathbf x\,\kappa\,\Bigl\{\hat R_2(\mathbf x),\frac{\delta\hat W}{\delta\tilde\pi(\mathbf x)}\Bigr\}+\int d^3\mathbf x\,i\mu\,\Bigl[\hat R_2(\mathbf x),\frac{\delta\hat W}{\delta\tilde h(\mathbf x)}\Bigr]~.\label{eq:disc_CQ_master_markov_final}
\end{align}
Here the local classical drift $\tilde A[\tilde h ,\tilde \pi]$ is
\begin{align}
\tilde A[\tilde h,\tilde\pi](\mathbf x)\equiv-\frac{\delta H_{\tilde h}}{\delta\tilde h(\mathbf x)}-\frac{D_{33}^{(1)}}{\hbar}\tilde\pi(\mathbf x)-\frac{D_{33}}{\hbar}\tilde h(\mathbf x),\label{eq:disc_markov_A_final}
\end{align}
the effective Hamiltonian acting on the $\psi$ sector is
\begin{align}
\hat H_{\rm eff}[\tilde h,\tilde\pi]=\hat H_\psi+\int d^3\mathbf x\,\Bigl(\lambda_1\,\tilde h(\mathbf x)-\frac{D_{23}}{\hbar}\tilde h(\mathbf x)-\frac{D_{23}^{(1)}}{\hbar}\tilde\pi(\mathbf x)\Bigr)\hat F_2(\mathbf x)+\frac{D_{22}}{2\hbar}\int d^3\mathbf x\,\hat F_2(\mathbf x)^2-\frac{D_{22}^{(1)}}{4\hbar}\int d^3\mathbf x\, \Bigl\{ \hat F_2(\mathbf x),\hat R_2(\mathbf x)\Bigr\}~,\label{eq:disc_markov_Heff_final}
\end{align}
and the hybrid coefficients are
\begin{align}
\gamma\equiv \frac{\lambda_1}{2}+\frac{D_{32}}{2\hbar},\qquad\nu\equiv \frac{2N_{23}}{\hbar}=\frac{2N_{32}}{\hbar},\qquad\kappa\equiv \frac{D_{32}^{(1)}}{2\hbar},\qquad\mu\equiv \frac{2N_{23}^{(2)}}{\hbar}=\frac{2N_{32}^{(2)}}{\hbar}.\label{eq:disc_markov_coeffs_final}
\end{align}
In deriving Eq.~\eqref{eq:disc_CQ_master_markov_final}, we reorganize the purely quantum sector into GKSL form,
\begin{align}
\mathcal L_{22}^{\rm GKSL}[\hat W]=\sum_{a,b=1}^2\int d^3\mathbf x\,(\mathsf D_0)_{ab}\left(\hat L_a(\mathbf x)\hat W\hat L_b(\mathbf x)-\frac12\bigl\{\hat L_b(\mathbf x)\hat L_a(\mathbf x),\hat W\bigr\}\right)~,\label{eq:disc_markov_GKSL22}
\end{align}
where $\mathsf D_0$ is the GKSL coefficient matrix, with components
\begin{align}
\mathsf D_0=\frac{2}{\hbar^2}
\begin{pmatrix}
N_{22} & -\frac{i}{4}D_{22}^{(1)}\\[2pt]
\frac{i}{4}D_{22}^{(1)} & N_{22}^{(2)}
\end{pmatrix}~,
\label{eq:disc_markov_D0_matrix}
\end{align}
in the basis
\begin{align}
\hat L_1(\mathbf x)\equiv \hat F_2(\mathbf x),\qquad\hat L_2(\mathbf x)\equiv \hat R_2(\mathbf x).
\end{align}

\subsection{Markovian effective action and complete positivity condition}

We reconstruct the local Markovian effective action from the time-local master equation obtained above.
The detailed derivation by Trotter decomposition~\cite{Trotter:1959ytf,Suzuki:1976be,Oppenheim:2023mox} is presented in Appendix~\ref{app:trotter}.
Using the result derived there, we obtain
\begin{align}
\exp\!\left[\frac{i}{\hbar}S_{{\rm eff},M}^{\rm W}\right]&\propto\exp\!\Bigg[-\frac12\int dt\int d^3\mathbf x\,\bigl(\bm a+\mathsf B_+\bm L_+ + \mathsf B_-\bm L_-\bigr)^T\mathsf Q\bigl(\bm a+\mathsf B_+\bm L_+ + \mathsf B_-\bm L_-\bigr)\Bigg]\nonumber\\
&\quad\times\exp\!\Bigg[\frac{i}{\hbar}S_{H,{\rm eff}}[\psi^+,\psi^-;\tilde h,\tilde\pi]+\int dt\int d^3\mathbf x\,\left(\bm L_+^{\,T}\mathsf D_0\,\bm L_--\frac12\,\bm L_+^{\,T}\mathsf D_0\,\bm L_+-\frac12\,\bm L_-^{\,T}\mathsf D_0^\ast\,\bm L_-\right)
\Bigg].\label{eq:disc_markov_SeffW_gauss}
\end{align}
Here
\begin{align}
\bm L_\pm(\mathbf x)\equiv
\begin{pmatrix}
F_2[\psi_\pm](\mathbf x)\\[2pt]
R_2[\psi_\pm](\mathbf x)
\end{pmatrix},
\qquad
\bm a\equiv
\begin{pmatrix}
\dot{\tilde h}-\tilde\pi\\[2pt]
-(\dot{\tilde\pi}-\tilde A)
\end{pmatrix},
\label{eq:disc_markov_L_a_def}
\end{align}
and
\begin{align}
\mathsf Q
\equiv
\begin{pmatrix}
(2N_{33}^{(2)})^{-1} & 0\\[2pt]
0 & (2N_{33})^{-1}
\end{pmatrix},
\qquad
\mathsf B_+
\equiv
\begin{pmatrix}
0 & -i\mu\\[2pt]
-(\gamma-i\nu) & -\kappa
\end{pmatrix},
\qquad
\mathsf B_-=\mathsf B_+^\ast .
\label{eq:disc_markov_QB_def}
\end{align}
The vector \(\bm a\) measures the deviation from the deterministic classical phase-space equations \(\dot{\tilde h}=\tilde\pi\) and \(\dot{\tilde\pi}=\tilde A\), while \(F_2[\psi_\pm]\) and \(R_2[\psi_\pm]\) in \(\bm L_\pm\) are the c-number branch variables corresponding to the operator insertions \(\hat F_2\) and \(\hat R_2\) in the path-integral representation.

The functional \(S_{H,{\rm eff}}[\psi^+,\psi^-;\tilde h,\tilde\pi]\) is the two-branch c-number action associated with the effective Hamiltonian \eqref{eq:disc_markov_Heff_final}.
More precisely, we define
\begin{align}
S_{H,{\rm eff}}[\psi^+,\psi^-;\tilde h,\tilde\pi]
\equiv
S_{H,{\rm eff}}[\psi^+;\tilde h,\tilde\pi]
-
S_{H,{\rm eff}}[\psi^-;\tilde h,\tilde\pi],
\label{eq:disc_markov_SHeff_two_branch}
\end{align}
with the single-branch action
\begin{align}
S_{H,{\rm eff}}[\psi;\tilde h,\tilde\pi]&\equiv S_0[\psi]+\int dt\int d^3\mathbf x\,\Biggl[\Bigl(\lambda_1\,\tilde h-\frac{D_{23}}{\hbar}\tilde h-\frac{D_{23}^{(1)}}{\hbar}\tilde\pi\Bigr)F_2[\psi]+\frac{D_{22}}{2\hbar}\,F_2[\psi]^2-\frac{D_{22}^{(1)}}{2\hbar}\,F_2[\psi]\,R_2[\psi]\Biggr].\label{eq:disc_markov_SHeff_branch}
\end{align}
This is simply the action representation of \(\hat H_{\rm eff}\), obtained by replacing the operator insertions \(\hat F_2\) and \(\hat R_2\) with the corresponding branch variables.

Expanding the exponent in Eq.~\eqref{eq:disc_markov_SeffW_gauss}, one finds the same structural separation as in Sec.~\ref{sec:bridge}: a purely classical part, a pair of single-branch classical--quantum couplings, and a genuine two-branch contribution.
We therefore write
\begin{align}
\frac{i}{\hbar}S_{{\rm eff},M}^{\rm W}=-I_C^{(M)}[\tilde h,\tilde\pi]+I_{CQ}^{(M)}[\tilde h,\tilde\pi,\psi_+]+\bigl(I_{CQ}^{(M)}[\tilde h,\tilde\pi,\psi_-]\bigr)^\ast+\tilde I_{CQ}^{(M)}[\tilde h,\tilde\pi,\psi_+,\psi_-].\label{eq:disc_markov_split_full_CQ}
\end{align}
The first term is the classical Gaussian weight associated with the phase-space residual \(\bm a\),
\begin{align}
I_C^{(M)}[\tilde h,\tilde\pi]=\frac12\int_{t_i}^{t_f}dt\int d^3\mathbf x\,\bm a^T\mathsf Q\,\bm a.\label{eq:disc_markov_IC}
\end{align}
It suppresses deviations from the deterministic classical equations of motion.

The second and third term collect all contributions that depend on a single branch.
Introducing
\begin{align}
\bm L[\psi]\equiv
\begin{pmatrix}
F_2[\psi]\\[2pt]
R_2[\psi]
\end{pmatrix},
\end{align}
it takes the form
\begin{align}
I_{CQ}^{(M)}[\tilde h,\tilde\pi,\psi]=\frac{i}{\hbar}S_{H,{\rm eff}}[\psi;\tilde h,\tilde\pi]-\int dt\int d^3\mathbf x\,\bm a^T\mathsf Q\,\mathsf B_+\bm L[\psi]-\frac12\int dt\int d^3\mathbf x\,\bm L[\psi]^T\bigl(\mathsf D_0+\mathsf B_+^{\,T}\mathsf Q\mathsf B_+\bigr)\bm L[\psi].\label{eq:disc_markov_ICQ}
\end{align}
The first term in Eq.~\eqref{eq:disc_markov_ICQ} is the Hamiltonian single-branch contribution inherited from \(\hat H_{\rm eff}\), while the remaining terms describe, respectively, the local hybrid coupling to the classical phase-space residual and the branch-diagonal quadratic term on the same branch.

The genuinely two-branch contribution comes from the mixed term in Eq.~\eqref{eq:disc_markov_SeffW_gauss}. A straightforward expansion gives
\begin{align}
\tilde I_{CQ}^{(M)}=\int dt\int d^3\mathbf x\,\bm L_+^{\,T}\Bigl(\mathsf D_0-\mathsf B_+^{\,T}\mathsf Q\mathsf B_-\Bigr)\bm L_-.\label{eq:disc_markov_Itilde_pre}
\end{align}
At this point it is convenient to introduce the two hybrid-coupling vectors
\begin{align}
\bm d_{\tilde\pi}=
\begin{pmatrix}
\gamma-i\nu\\[2pt]
\kappa
\end{pmatrix},
\qquad
\bm d_{\tilde h}=
\begin{pmatrix}
0\\[2pt]
i\mu
\end{pmatrix},
\label{eq:disc_markov_blocks}
\end{align}
in terms of which
\begin{align}
\mathsf B_+
=
-
\begin{pmatrix}
\bm d_{\tilde h}^{\,T}\\[2pt]
\bm d_{\tilde\pi}^{\,T}
\end{pmatrix},
\qquad
\mathsf B_-
=
-
\begin{pmatrix}
(\bm d_{\tilde h}^{\,\dagger})^{T}\\[2pt]
(\bm d_{\tilde\pi}^{\,\dagger})^{T}
\end{pmatrix}.
\label{eq:disc_markov_B_from_d}
\end{align}
The mixed term can then be written more transparently as
\begin{align}
\tilde I_{CQ}^{(M)}=\int dt \int d^3\mathbf x\, d^3\mathbf y\,L_{+,a}(\mathbf x)\,C_M^{ab}(\mathbf x,\mathbf y)\,L_{-,b}(\mathbf y),\label{eq:disc_markov_Itilde}
\end{align}
where the local kernel is
\begin{align}
C_M^{ab}(\mathbf x,\mathbf y)=(\mathsf C_M)_{ab}\,\delta^{(3)}(\mathbf x-\mathbf y),\label{eq:disc_markov_CM_kernel}
\end{align}
with
\begin{align}
\mathsf C_M&=\mathsf D_0-\mathsf B_+^{\,T}\mathsf Q\mathsf B_-\nonumber\\
&=\mathsf D_0-\frac{1}{2N_{33}^{(2)}}\,\bm d_{\tilde h}\bm d_{\tilde h}^{\,\dagger}-\frac{1}{2N_{33}}\,\bm d_{\tilde\pi}\bm d_{\tilde\pi}^{\,\dagger}.\label{eq:disc_markov_CM}
\end{align}
This kernel is the local Markovian analogue of the nonlocal kernel \(C(x,y)\) that controlled the non-Markovian positivity condition.

The complete-positivity condition now follows in the same way as in Sec.~\ref{sec:bridge}.
If \(\mathsf C_M\) is positive semidefinite, then the branch-coupling factor admits the complex Gaussian unraveling
\begin{align}
e^{\tilde I_{CQ}^{(M)}}=\int \mathcal D\eta\,\mathcal D\eta^\ast\,P_M[\eta]\,\exp\!\left[i\int dt\int d^3\mathbf x\,\eta^T \bm L_+-i\int dt\int d^3\mathbf x\,\eta^\dagger \bm L_-\right],\label{eq:disc_markov_unravel}
\end{align}
with
\begin{align}
P_M[\eta]\propto\exp\!\left[-\int dt\int d^3\mathbf x\,\eta^\dagger \mathsf C_M^{-1}\eta\right].\label{eq:disc_markov_eta_weight}
\end{align}
Here \(\eta\) is a two-component complex auxiliary field; \(\eta^T\) denotes transpose, and \(\eta^\dagger\) its Hermitian conjugate.
Therefore the positivity of \(\mathsf C_M\), equivalently
\begin{align}
C_M\succeq 0,\label{eq:disc_markov_CP_CM_kernel}
\end{align}
is a sufficient condition for representing the two-branch term as an average over completely positive single-noise realizations.
In addition, the conditions
\begin{align}
N_{33}\ge 0,
\qquad
N_{33}^{(2)}\ge 0,
\label{eq:disc_markov_CP_classical}
\end{align}
ensure that Eq.~\eqref{eq:disc_markov_IC} defines a well-defined probability measure.
Together, Eq.~\eqref{eq:disc_markov_CP_CM_kernel} and Eq.~\eqref{eq:disc_markov_CP_classical} provide conditions for positivity preservation of the semi-Wigner operator under the Markovian evolution.

\subsection{Dictionary to the Oppenheim model and the CQ trade-off}
We now compare the time-local equation \eqref{eq:disc_CQ_master_markov_final} with the Markovian classical-quantum master equation reviewed in Appendix~\ref{sec:CQreview}. 
The purpose of this subsection is to make explicit the dictionary between the coefficients appearing in our Markovian hidden-model description and those of the Oppenheim classical--quantum master equation.

At this stage we identify the semi-Wigner operator with the CQ state,
\begin{align}
\hat\rho_{\rm CQ}(z)\equiv \hat W[\tilde h,\tilde\pi]~,\label{eq:disc_markov_identification}
\end{align}
where \(z=(\tilde h,\tilde\pi)\) denotes the classical phase-space point.
With this identification, Eq.~\eqref{eq:disc_CQ_master_markov_final} takes the Oppenheim form, with ordinary derivatives with respect to the classical phase-space variables replaced by functional derivatives.

We begin with the purely classical sector.
The drift coefficients are read off directly from the Liouville part of the master equation:
\begin{align}
D_{1,\tilde h(\mathbf x)}^{00}(z)=\tilde\pi(\mathbf x),\qquad D_{1,\tilde\pi(\mathbf x)}^{00}(z)=\tilde A[\tilde h,\tilde\pi](\mathbf x).\label{eq:disc_markov_drift_dict}
\end{align}
Similarly, the nonvanishing components of the classical diffusion block are
\begin{align}
D_{2,\tilde h(\mathbf x)\tilde h(\mathbf y)}^{00}(z)=N_{33}^{(2)}\delta^{(3)}(\mathbf x-\mathbf y),\qquad D_{2,\tilde\pi(\mathbf x)\tilde\pi(\mathbf y)}^{00}(z)=N_{33}\delta^{(3)}(\mathbf x-\mathbf y).
\label{eq:disc_markov_D200_dict}
\end{align}

We next turn to the purely quantum sector.
Choosing the Lindblad basis
\begin{align}
\alpha=(a,\mathbf x),\qquad\beta=(b,\mathbf y),\qquad a,b\in\{1,2\},\label{eq:disc_markov_L_basis_indices}
\end{align}
with
\begin{align}
\hat L_{1,\mathbf x}=\hat F_2(\mathbf x),\qquad\hat L_{2,\mathbf x}=\hat R_2(\mathbf x),\label{eq:disc_markov_L_basis}
\end{align}
we find that the purely quantum GKSL block is
\begin{align}
D_0^{\alpha\beta}(z)
=
(\mathsf D_0)_{ab}\,
\delta^{(3)}(\mathbf x-\mathbf y),
\label{eq:disc_markov_D0_dict}
\end{align}
where the \(2\times2\) matrix \(\mathsf D_0\) is given in Eq.~\eqref{eq:disc_markov_D0_matrix}.

Finally, the coupling between the classical and quantum sectors is encoded in the hybrid coefficients.
At the order retained in the Markov expansion, the only nonvanishing components are
\begin{align}
D_{1,\tilde\pi(\mathbf x)}^{0(a,\mathbf x)}(z)
&=(\bm d_{\tilde\pi})_a\,,\qquad D_{1,\tilde\pi(\mathbf x)}^{(a,\mathbf x)0}(z)=(\bm d_{\tilde\pi})_a^\ast\,,\label{eq:disc_markov_D1pi_dict}\\
D_{1,\tilde h(\mathbf x)}^{0(a,\mathbf x)}(z)&=(\bm d_{\tilde h})_a\,,\qquad D_{1,\tilde h(\mathbf x)}^{(a,\mathbf x)0}(z)=(\bm d_{\tilde h})_a^\ast\,.\label{eq:disc_markov_D1h_dict}
\end{align}
All remaining hybrid coefficients vanish at this order.
This completes the dictionary between the Markovian hidden model and the Oppenheim classical--quantum model.

Under this dictionary, the local kernel~\eqref{eq:disc_markov_CM_kernel} is written as
\begin{align}
C_M=D_0(z)-D_{1,i}^0(z)\bigl(2D_{2,ij}^{00}(z)\bigr)^{-1}D_{1,j}^0(z)^\dagger,\label{eq:disc_markov_C_schur}
\end{align}
where the inverse is understood as the generalized inverse on the support of $D_2^{00}$.
Hence, whenever $D_0(z)$ is invertible on its support, the hidden-model positivity condition~\eqref{eq:disc_markov_CP_CM_kernel} is equivalent to
\begin{align}
2D_2^{00}(z)\succeq D_1^{0\alpha}(z)\,D_0^{\alpha\beta}(z)^{-1}\,D_1^{0\beta}(z)^\dagger.\label{eq:disc_markov_tradeoff}
\end{align}
This is precisely the CQ trade-off condition reviewed in Appendix~\ref{sec:CQreview}.

The main result of this subsection is that the Markovian hidden model reproduces a subclass of the Markovian classical--quantum dynamics considered by Oppenheim, which may also be viewed from the perspective of a Stinespring-type dilation~\cite{Hotta:2025eew}.
In the short memory regime, the classical drift and diffusion, the quantum GKSL block, and the hybrid couplings are generated by the local moment expansion of the underlying influence kernels.
This suggests that experimental agreement with such classical--quantum dynamics does not by itself imply that the underlying theory is fundamentally classical--quantum.
Instead, it may equally well admit a hidden quantum origin.

\section{Summary}\label{sec:discussion}

In this work, we showed that classical--quantum dynamics arise generically as an effective description of fully quantum systems under decoherence, characterized by a positivity condition on the underlying nonlocal kernels.
To this end, we constructed a hidden model of interacting scalar fields and derived the reduced dynamics by tracing out an unobserved sector.
Using the semi-Wigner representation, we identified a regime within the resulting non-Markovian dynamics in which one field admits an effective classical interpretation.
We further showed that, in the Markovian case, the reduced dynamics reproduce a subclass of the classical--quantum models proposed by Oppenheim and collaborators~\cite{Oppenheim:2018igd,Oppenheim:2022xjc,Oppenheim:2023mox,Carney:2024izr}.

Compared with conventional classical--quantum formulations, the results clarify two key structural features.
First, a fundamentally classical sector need not be postulated from the outset.
Instead, the hybrid dynamics can be derived systematically from an underlying quantum theory with decoherence.
Second, the present construction is naturally formulated at the non-Markovian level.
More broadly, this framework provides a systematic route to deriving effective classical--quantum dynamics from microscopic quantum models, both in gravitational settings and beyond.

Our explicit analysis was carried out in a scalar-field model, mainly to avoid the additional complications associated with the gauge structure of gravity.
We therefore do not claim to have derived an effective classical--quantum description of gravity itself.
Nevertheless, the present results suggest that care is needed when interpreting proposed tests of the quantum nature of gravity, including BMV-type scenarios~\cite{Bose:2017nin,Marletto:2017kzi} and tests based on classical--quantum models~\cite{Miki:2025wqf,Fabiano:2026agi}.
Even if an experiment is found to be consistent with classical--quantum dynamics at the level of reduced observables, this would not by itself establish that the mediator is fundamentally classical, because similar dynamics may arise effectively within an underlying fully quantum theory.
Conversely, ruling out effective classical--quantum descriptions of this kind requires probing regimes that cannot be embedded in the present hidden-model framework.
A representative example is dynamics with genuinely nonlinear dependence on the density matrix, such as in the Schr\"odinger--Newton equation.
Such dynamics cannot be obtained here, because the effective evolution always descends from linear quantum dynamics followed by an environmental trace.

Taken together, our results sharpen a broader conceptual question:
to what extent can one experimentally distinguish a theory in which classical--quantum structure is fundamental from a hidden quantum model in which the same classical--quantum dynamics arise only effectively through decoherence?
Clarifying this distinction will be useful for interpreting future experiments on the quantum or classical nature of gravity, and more generally on the status of mediator fields.

\section*{Acknowledgements}
We would like to thank Takahiro Tanaka for useful comments.
S.~T. is supported by the Research Fellow program of Kyoto University.
H.~T. is supported by the Hakubi project at Kyoto University and by Japan Society for the Promotion of Science (JSPS) KAKENHI Grant No.~JP22K14037.

\appendix
\section{Cubic Interaction Model as an Illustrative Example}\label{sec:example}

In this section, we illustrate the general framework by applying it to a concrete model with explicit cubic interactions, where integrating out an environmental field generates non-Markovian kernels in the effective dynamics.
This model provides a concrete realization of the hidden model introduced in Sec.~\ref{sec:bridge}.

We consider three scalar fields: $h$, $\psi$, and $\phi$.
The field $h$ plays the role of an effective mediator, $\psi$ represents the quantum matter sector, and $\phi$ is an unobserved environment.
Motivated by the structure of linearized gravity, we consider interactions that are linear in $h$ and quadratic in the matter fields:
\begin{align}
    S_\mathrm{int}[h,\psi] &= \int d^4x\, \lambda_1\, h\,\psi^2~,\quad
    S_\mathrm{int}[\psi,\phi] = \int d^4x\, \lambda_2\, \psi^2\,\phi~,\quad
    S_\mathrm{int}[h,\phi] = \int d^4x\, \lambda_3\, h\,\phi^2~,
\end{align}
with the free action of $h$
\begin{align}
    S_0[h]=\int d^4x \left[\frac{(\partial h)^2}{2}-\frac{m_\mathrm{bare}^2}{2}h^2\right]~.
\end{align}
In the notation of Sec.~\ref{sec:Effac}, this corresponds to choosing $F_2[\psi]=\psi^2$ and $F_3[h]=h$ (hence $\tilde h=h$), with $G_2[\phi]=\phi$ and $G_3[\phi]=\phi^2$.

With these interactions, the effective action for the semi-Wigner operator,
defined in Eq.~\eqref{eq:SeffW_from_delta}, takes the form
\begin{align}
     \exp\!\left[\frac{i}{\hbar}S_\mathrm{eff}^\mathrm{W}[\psi^+,\psi^-,h^c]\right]
     &=\delta(\pi^c_i-\dot{h}_i^c)\delta(\pi_f^c-\dot{h}_f^c)
     \int \mathcal{D}\xi\,\delta\!\left(\ddot{h}^c-(\nabla^2-m^2)h^c-\lambda_1\Psi_c^2
     +\frac{1}{2\hbar}\!\int\! d^4y\,\mathcal{D}_h(x,y)h^c(y)-\xi(x)\right)\nonumber\\
    &\quad\times\exp\!\bigg[-\frac{1}{2}\!\int\! d^4x\,d^4y\,\xi(x)\mathcal{N}_h^{-1}(x,y)\xi(y)
    +\frac{i}{\hbar}\!\int\! d^4x\,\lambda_1h^c(x)\Psi_\Delta^2(x)
    +\frac{i}{\hbar}\big(S_0[\psi^+]-S_0[\psi^-]\big)\bigg]\nonumber\\
    &\quad\times\exp\!\bigg[-\frac{1}{2\hbar^2}\!\int\! d^4x\,d^4y\,
    \bigg(\Psi_\Delta^2(x)\mathcal{N}_\psi(x,y)\Psi_\Delta^2(y)
    +i\Psi_\Delta^2(x)\mathcal{D}_\psi(x,y)\Psi_c^2(y)\bigg)\bigg]~.\label{eq:ouraction}
\end{align}
Here
\begin{align}
    \Psi_\Delta^2(x)&\equiv \psi_+^2(x)-\psi_-^2(x)~,\\
    \Psi_c^2(x)&\equiv \tfrac{1}{2}\big(\psi_+^2+\psi_-^2\big)~,
\end{align}
and the noise and dissipation kernels are 
\begin{align}
    \mathcal{N}_\psi(x,y)&=4\lambda_2^2\,\Re[G_F(x,y)]~, \quad\mathcal{D}_\psi(x,y)=4\lambda_2^2\,\theta(x^0-y^0)\Im[G_F(x,y)]~,\\
    \mathcal{N}_h(x,y)&=8\lambda_3^2\,\Re[G_F(x,y)^2]~, \quad 
    \mathcal{D}_h(x,y)=8\lambda_3^2\,\theta(x^0-y^0)\Im[G_F(x,y)^2]~,
\end{align}
with $G_F(x,y)$ is the Feynman propagator of the environmental field $\phi$.  
In addition, for an initial state that is symmetric under $\phi\to-\phi$ (for example, a Gaussian state with vanishing odd moments), the mixed correlators $\langle \phi(x)\phi^2(y)\rangle$ vanish.
Consequently, the mixed kernels $\mathcal N_{23},\mathcal D_{23}$ and $\mathcal N_{32},\mathcal D_{32}$ introduced in Sec.~\ref{sec:Effac} also vanish.
For this reason, the effective action takes the relatively simple form in the present illustrative model.

Applying the general decomposition~\eqref{eq:split_full_CQ}, the effective action takes the form~\eqref{eq:split_full_CQ}.
The purely classical part, corresponding to Eq.~\eqref{eq:IC_def_section}, is given by the $h^c$-dependent part of the Langevin kernel as
\begin{align}
    I_C[h^c]\equiv \frac12\int d^4x\,d^4y~\mathcal{E}_h[h^c](x)\,\mathcal N_h^{-1}(x,y)\,\mathcal{E}_h[h^c](y),\label{eq:IC_example_new}
\end{align}
with
\begin{align}
    \mathcal{E}_h[h^c](x)\equiv\ddot{h}^c(x)-(\nabla^2-m^2)h^c(x) +\frac{1}{2\hbar}\!\int\! d^4y\,\mathcal{D}_h(x,y)h^c(y).
\end{align}
The single branch contribution is obtained from Eq.~\eqref{eq:ICQ_explicit} as
\begin{align}
    I_{CQ}[h,\psi]&\equiv \frac{\lambda_1}{2}\int d^4x\,d^4y~\psi(x)^2\mathcal{N}_h^{-1}(x,y)\bigg(\ddot{h}^c(y)-(\nabla^2-m^2)h^c(y) +\frac{1}{2\hbar}\int d^4z\,\mathcal{D}_h(y,z)\,h^c(z) \bigg)\nonumber\\
    &\hspace{10mm}-\int d^4x\,d^4y~\psi(x)^2\bigg( \frac{1}{2\hbar^2}\mathcal{N}_\psi(x,y)+\frac{\lambda_1^2}{8}\mathcal{N}_h^{-1}(x,y)\bigg)\psi(y)^2+\frac{i\lambda_1}{\hbar}\int d^4x\,h^c(x)\psi(x)^2\nonumber\\
    &\hspace{15mm}-\frac{i}{4\hbar^2}\int d^4x\,d^4y~\psi(x)^2\mathcal{D}_\psi(x,y)\psi(y)^2
    +\frac{i}{\hbar}S_0[\psi]~.
\end{align}
The remaining coupling between two branches can be written as the bilinear form,
\begin{align}
    \widetilde I_{CQ}[h^c,\psi_+,\psi_-]=\int d^4x\,d^4y~\psi_+^2(x)\,C(x,y)\,\psi_-^2(y)~. \label{eq:Itilde_example_new}
\end{align}
 The corresponding kernel is obtained from the general formula~\eqref{eq:C_compact_section} as
\begin{align}
    C=\frac{1}{\hbar^2}\mathcal N_\psi-\frac{\lambda_1^2}{4}\,\mathcal N_h^{-1}-\frac{i}{2\hbar^2}\,\mathcal D_\psi^{a},\qquad\mathcal D_\psi^{a}(x,y)\equiv \frac12\Big(\mathcal D_\psi(x,y)-\mathcal D_\psi(y,x)\Big)~.\label{eq:C_example_new}
\end{align}
Applying the sufficient condition~\eqref{eq:CP_C_section} to the kernel~\eqref{eq:C_example_new} ensures complete positivity.
This does not imply CP divisibility, nor does it imply that the instantaneous generator admits a GKSL form away from the Markov limit.

We now specialize the general Markovian dictionary of Sec.~\ref{sec:markov} to the present cubic model.
In the local white-noise approximation, we take
\begin{align}
    \mathcal N_\psi(x,y)\simeq 2N_{22}\,\delta^{(4)}(x-y),\qquad
    \mathcal N_h(x,y)\simeq 2N_{33}\,\delta^{(4)}(x-y),\qquad
    \mathcal D_\psi\simeq 0,\qquad
    \mathcal D_h\simeq 0~.
    \label{eq:example_white_noise_ident}
\end{align}
Because $F_2[\psi]=\psi^2$in this model, the corresponding operators are
\begin{align}
    \hat F_2(\mathbf x)=\hat\psi(\mathbf x)^2,\qquad
    \hat R_2(\mathbf x)=\frac{i}{\hbar}[\hat H_\psi,\hat F_2(\mathbf x)].
\end{align}
Furthermore, since the mixed kernels vanish, the hybrid coefficients simplify to
\begin{align}
    \nu=0,\qquad \kappa=0,\qquad \gamma=\frac{\lambda_1}{2}.
\end{align}
The general trade-off condition \eqref{eq:disc_markov_tradeoff} therefore becomes
\begin{align}
    2N_{33}\ \ge\ \frac{\bigl(\lambda_1/2\bigr)^2}{\,2N_{22}/\hbar^2\,}
    \qquad\Longleftrightarrow\qquad
    N_{22}N_{33}\ge \frac{\hbar^2\lambda_1^2}{16},
    \label{eq:example_tradeoff_N22N33}
\end{align}
in agreement with the model used in \cite{Carney:2024izr}.

\section{Trotter reconstruction of the local Markovian effective action}
\label{app:trotter}

In this appendix, we derive the local Markovian effective action~\eqref{eq:disc_markov_SeffW_gauss} corresponding to the time-local master equation
\eqref{eq:disc_CQ_master_markov_final}.
Throughout this appendix, we suppress the spatial coordinate $\mathbf x$ in the intermediate steps.
The full field-theoretic expression is recovered at the end by restoring $\int dt\,d^3\mathbf x$ together with the locality factor $\delta^{(3)}(\mathbf x-\mathbf y)$.

Let
\begin{align}
\partial_t\hat W=\mathcal L_M\hat W~,
\end{align}
denote the Markovian master equation \eqref{eq:disc_CQ_master_markov_final}.
We split the generator as
\begin{align}
\mathcal L_M=\mathcal L_{\rm cl}-\frac{i}{\hbar}\bigl[\hat H_{\rm eff}[\tilde h,\tilde\pi],\,\cdot\,\bigr]+\mathcal L_{22}^{\rm GKSL}+\mathcal L_{\rm hyb},
\end{align}
where
\begin{align}
\mathcal L_{\rm cl}\hat W&=-\tilde\pi\,\frac{\partial \hat W}{\partial \tilde h}-\tilde A[\tilde h,\tilde\pi]\,\frac{\partial \hat W}{\partial \tilde\pi}+N_{33}\,\frac{\partial^2 \hat W}{\partial \tilde\pi^2}+N_{33}^{(2)}\,\frac{\partial^2 \hat W}{\partial \tilde h^2},\\
\mathcal L_{\rm hyb}\hat W&=\gamma\Bigl\{\hat F_2,\frac{\partial\hat W}{\partial\tilde\pi}\Bigr\}-i\nu\Bigl[\hat F_2,\frac{\partial\hat W}{\partial\tilde\pi}\Bigr]+\kappa\Bigl\{\hat R_2,\frac{\partial\hat W}{\partial\tilde\pi}\Bigr\}+i\mu\Bigl[\hat R_2,\frac{\partial\hat W}{\partial\tilde h}\Bigr].
\end{align}
The short-time kernel for a time step $\epsilon$ is
\begin{align}
\hat W_{n+1}(\tilde h',\tilde\pi')=\int d\tilde h\,d\tilde\pi\;\mathcal J_\epsilon(\tilde h',\tilde\pi'|\tilde h,\tilde\pi)\,\hat W_n(\tilde h,\tilde\pi),
\end{align}
with
\begin{align}
\mathcal J_\epsilon(\tilde h',\tilde\pi'|\tilde h,\tilde\pi)=\Bigl(1+\epsilon\,\mathcal L_M^{(\tilde h',\tilde\pi')}\Bigr)\delta(\tilde h'-\tilde h)\,\delta(\tilde\pi'-\tilde\pi)+\mathcal O(\epsilon^2).\label{eq:markov_short_kernel_start}
\end{align}
We represent the phase-space delta functions by
\begin{align}
\delta(\tilde h'-\tilde h)\,\delta(\tilde\pi'-\tilde\pi)=\int\frac{d\tilde\pi_\Delta\,d\tilde h_\Delta}{(2\pi\hbar)^2}\exp\!\left[\frac{i}{\hbar}\tilde\pi_\Delta(\tilde h'-\tilde h)-\frac{i}{\hbar}\tilde h_\Delta(\tilde\pi'-\tilde\pi)\right].\label{eq:markov_delta_rep}
\end{align}
Since the derivatives in \eqref{eq:markov_short_kernel_start} act on the final variables,
\begin{align}
\frac{\partial}{\partial\tilde h'}\rightarrow \frac{i}{\hbar}\tilde\pi_\Delta,\qquad\frac{\partial}{\partial\tilde\pi'}\rightarrow -\frac{i}{\hbar}\tilde h_\Delta.\label{eq:markov_derivative_rules}
\end{align}
Substituting \eqref{eq:markov_delta_rep} and \eqref{eq:markov_derivative_rules} into \eqref{eq:markov_short_kernel_start}, and keeping terms through $\mathcal O(\epsilon)$, we obtain
\begin{align}
\mathcal J_\epsilon&\propto\int\frac{d\tilde\pi_\Delta\,d\tilde h_\Delta}{(2\pi\hbar)^2}\exp\!\left[\frac{i}{\hbar}\tilde\pi_\Delta\bigl(\Delta\tilde h-\epsilon\tilde\pi\bigr)-\frac{i}{\hbar}\tilde h_\Delta\bigl(\Delta\tilde\pi-\epsilon\tilde A\bigr)-\epsilon\frac{N_{33}^{(2)}}{\hbar^2}\tilde\pi_\Delta^{\,2}-\epsilon\frac{N_{33}}{\hbar^2}\tilde h_\Delta^{\,2}\right]\nonumber\\
&\qquad\times\exp\!\left[-\frac{i\epsilon}{\hbar}\hat H_{{\rm eff},+}+\frac{i\epsilon}{\hbar}\hat H_{{\rm eff},-}+\epsilon\,\bm{\hat L}_+^{\,T}\mathsf D_0\,\bm{\hat L}_--\frac{\epsilon}{2}\bm{\hat L}_+^{\,T}\mathsf D_0\,\bm{\hat L}_+-\frac{\epsilon}{2}\bm{\hat L}_-^{\,T}\mathsf D_0^{\ast}\,\bm{\hat L}_-\right]\nonumber\\
&\qquad\times\exp\!\left[-\frac{i\epsilon}{\hbar}\tilde h_\Delta\Bigl((\gamma-i\nu)\hat F_{2,+}+(\gamma+i\nu)\hat F_{2,-}+\kappa \hat R_{2,+}+\kappa \hat R_{2,-}\Bigr)-\frac{i\epsilon}{\hbar}\tilde\pi_\Delta\Bigl(-i\mu \hat R_{2,+}+i\mu \hat R_{2,-}\Bigr)\right],\label{eq:markov_short_kernel_full}
\end{align}
where
\begin{align}
\Delta\tilde h\equiv \tilde h_{n+1}-\tilde h_n,\qquad\Delta\tilde\pi\equiv \tilde\pi_{n+1}-\tilde\pi_n,\qquad\bm{\hat L}_\pm\equiv
\begin{pmatrix}
\hat F_{2,\pm}\\
\hat R_{2,\pm}
\end{pmatrix}.
\end{align}
Passing to the continuum limit and restoring the branch path integrals for the $\psi$ sector, we obtain
\begin{align}
\exp\!\left[\frac{i}{\hbar}S_{{\rm eff},M}^{\rm W}[\psi^+,\psi^-,\tilde h,\tilde\pi]\right]&\propto\int \mathcal D\tilde\pi_\Delta\,\mathcal D\tilde h_\Delta\;\exp\!\left[-\int dt\,d^3\mathbf x\,\left(\frac{N_{33}^{(2)}}{\hbar^2}\tilde\pi_\Delta^{\,2}+\frac{N_{33}}{\hbar^2}\tilde h_\Delta^{\,2}\right)\right]\nonumber\\
&\quad\times\exp\!\left[\frac{i}{\hbar}\int dt\,d^3\mathbf x\,\Bigl\{\tilde\pi_\Delta\,(\dot{\tilde h}-\tilde\pi)-\tilde h_\Delta\,(\dot{\tilde\pi}-\tilde A)\Bigr\}\right]\nonumber\\
&\quad\times\exp\!\left[-\frac{i}{\hbar}\int dt\,d^3\mathbf x\,\tilde h_\Delta\Bigl((\gamma-i\nu)F_2[\psi_+]+(\gamma+i\nu)F_2[\psi_-]+\kappa R_2[\psi_+]+\kappa R_2[\psi_-]\Bigr)\right]\nonumber\\
&\quad\times\exp\!\left[-\frac{i}{\hbar}\int dt\,d^3\mathbf x\,\tilde\pi_\Delta\Bigl(-i\mu\,R_2[\psi_+]+i\mu\,R_2[\psi_-]\Bigr)\right]\,e^{\frac{i}{\hbar}S_{H,{\rm eff}}[\psi^+,\psi^-;\tilde h,\tilde\pi]}\nonumber\\
&\quad\times\exp\!\left[\int dt\,d^3\mathbf x\,\left(\bm L_+^{\,T}\mathsf D_0\,\bm L_--\frac12\bm L_+^{\,T}\mathsf D_0\,\bm L_+-\frac12\bm L_-^{\,T}\mathsf D_0^{\ast}\,\bm L_-\right)\right],\label{eq:markov_SeffW_response}
\end{align}
where
\begin{align}
\bm L_\pm
\equiv
\begin{pmatrix}
F_2[\psi_\pm]\\
R_2[\psi_\pm]
\end{pmatrix},
\end{align}
and \(S_{H,{\rm eff}}[\psi^+,\psi^-;\tilde h,\tilde\pi]\) denotes the contributions generated by the effective Hamiltonian \eqref{eq:disc_markov_Heff_final}.

It is convenient to introduce the two-component vectors
\begin{align}
\bm a
&\equiv
\begin{pmatrix}
\dot{\tilde h}-\tilde\pi\\[2pt]
-(\dot{\tilde\pi}-\tilde A)
\end{pmatrix},
\qquad
\bm v
\equiv
\begin{pmatrix}
\tilde\pi_\Delta\\[2pt]
\tilde h_\Delta
\end{pmatrix},
\\
\bm d_{\tilde h}
&\equiv
\begin{pmatrix}
0\\
i\mu
\end{pmatrix},
\qquad
\bm d_{\tilde\pi}
\equiv
\begin{pmatrix}
\gamma-i\nu\\
\kappa
\end{pmatrix},
\qquad
\mathsf Q
\equiv
\begin{pmatrix}
(2N_{33}^{(2)})^{-1} & 0\\
0 & (2N_{33})^{-1}
\end{pmatrix},
\end{align}
together with
\begin{align}
\mathsf B_+
\equiv
\begin{pmatrix}
0 & -i\mu\\
-(\gamma-i\nu) & -\kappa
\end{pmatrix},
\qquad
\mathsf B_-\equiv \mathsf B_+^\ast
=
\begin{pmatrix}
0 & i\mu\\
-(\gamma+i\nu) & -\kappa
\end{pmatrix}.
\end{align}
Then Eq.~\eqref{eq:markov_SeffW_response} becomes
\begin{align}
\exp\!\left[\frac{i}{\hbar}S_{{\rm eff},M}^{\rm W}\right]&\propto\int \mathcal D\bm v\;\exp\!\left[-\int dt\,d^3\mathbf x\,\bm v^{\,T}\mathsf Q^{-1}\bm v+\frac{i}{\hbar}\int dt\,d^3\mathbf x\,\bm v^{\,T}\bigl(\bm a+\mathsf B_+\bm L_+ + \mathsf B_-\bm L_-\bigr)\right]\nonumber\\
&\qquad\times e^{\frac{i}{\hbar}S_{H,{\rm eff}}[\psi^+,\psi^-;\tilde h,\tilde\pi]}\exp\!\left[\int dt\,d^3\mathbf x\,\left(\bm L_+^{\,T}\mathsf D_0\,\bm L_--\frac12\bm L_+^{\,T}\mathsf D_0\,\bm L_+-\frac12\bm L_-^{\,T}\mathsf D_0^{\ast}\,\bm L_-\right)\right].
\end{align}
The Gaussian integral over \(\bm v\) is elementary and gives
\begin{align}
\exp\!\left[\frac{i}{\hbar}S_{{\rm eff},M}^{\rm W}\right]&\propto\exp\!\left[-\frac12\int dt\,d^3\mathbf x\,\bigl(\bm a+\mathsf B_+\bm L_+ + \mathsf B_-\bm L_-\bigr)^{T}\mathsf Q\bigl(\bm a+\mathsf B_+\bm L_+ + \mathsf B_-\bm L_-\bigr)\right]\nonumber\\
&\qquad\times e^{\frac{i}{\hbar}S_{H,{\rm eff}}[\psi^+,\psi^-;\tilde h,\tilde\pi]}\exp\!\left[\int dt\,d^3\mathbf x\,\left(\bm L_+^{\,T}\mathsf D_0\,\bm L_--\frac12\bm L_+^{\,T}\mathsf D_0\,\bm L_+-\frac12\bm L_-^{\,T}\mathsf D_0^{\ast}\,\bm L_-\right)\right].\label{eq:markov_SeffW_gauss}
\end{align}

\section{Brief Review of the Classical–Quantum Framework}\label{sec:CQreview}

In this section, we briefly review the definition of a classical-quantum (CQ) state and the corresponding CQ action, following the formulation of Oppenheim and collaborators~\cite{Oppenheim:2023mox,Oppenheim:2018igd,Oppenheim:2022xjc}.  
We denote the classical field and its conjugate momentum by $(h,\pi)$ and the quantum field by $\psi$.

In this framework, the state of a classical–quantum system is encoded in an operator-valued function 
$\hat{\rho}_\mathrm{CQ}(h,\pi)$, which may be regarded as an unnormalized quantum operator assigned to each point in the classical phase space.  
From this object, one recovers both the full quantum state and the classical probability distribution,
\begin{align}
    &\hat{\rho}=\int dh\, d\pi\, \hat{\rho}_\mathrm{CQ}(h,\pi)~,\\
    &p(h,\pi)=\mathrm{Tr}\!\left[\hat{\rho}_\mathrm{CQ}(h,\pi)\right]~.
\end{align}
Thus, $\hat{\rho}_\mathrm{CQ}$ encodes all statistical information about the quantum and classical sectors and their mutual correlations.

The time evolution of such a state may be written formally in path-integral form as
\begin{align}
    \bra{\psi^+_f}\hat{\rho}_\mathrm{CQ}(h_f,\pi_f;t_f)\ket{\psi^-_f}
    =\!\int\! \mathcal{D}\psi^+\mathcal{D}\psi^-\mathcal{D}h\,\mathcal{D}\pi\,
    e^{I_\mathrm{CQ}[\psi^+,\psi^-,h,\pi]}\,
    \bra{\psi_i^+}\hat{\rho}_\mathrm{CQ}(h_i,\pi_i;t_i)\ket{\psi^-_i}~,\label{eq:CQ_Opp}
\end{align}
where $I_\mathrm{CQ}$ is the CQ action governing the coupled dynamics.  
Alternatively, in the quantum-information formulation one may write
\begin{align}
    \hat{\rho}_\mathrm{CQ}(z_f;t_f)
    =\!\int\! dz_i \sum_\mu
    \hat{K}_\mu(z_f|z_i)\,
    \hat{\rho}_\mathrm{CQ}(z_i;t_i)\,
    \hat{K}_\mu^\dagger(z_f|z_i)~,
\end{align}
where $z=(h,\pi)$.  
The Kraus operators $\hat{K}_\mu$ define a completely positive Markovian evolution provided the decoherence diffusion trade-off condition is satisfied.

Once the Kraus operators $\hat K_\mu$ specifying a
Markovian CQ evolution are given, expanding the map to first order
in infinitesimal time $\delta t$ yields a time-local master equation for $\hat\rho_{\rm CQ}(z;t)$.
In particular, Oppenheim \emph{et al.} showed that complete positivity imposes a CQ analogue of the Pawula theorem:
for a non-trivial CQ evolution, the generator must either involve an infinite Kramers--Moyal hierarchy(i.e.\ infinitely many non-vanishing moments of the CQ transition kernel), or else it truncates at second order.
In the latter case, the Markovian master equation takes the form~\cite{Oppenheim:2023mox}
\begin{align}
\frac{\partial \hat{\rho}_\mathrm{CQ}(z;t)}{\partial t}=&\;\sum_{n=1}^{2}(-1)^n\left(\frac{\partial^n}{\partial z_{i_1}\cdots \partial z_{i_n}}\right)\Big(D^{00}_{n,i_1\cdots i_n}(z)\,\hat{\rho}_\mathrm{CQ}(z;t)\Big)+\frac{\partial}{\partial z_i}\Big(D^{0\alpha}_{1,i}(z)\,\hat{\rho}_\mathrm{CQ}(z;t)\,\hat L_\alpha^\dagger\Big)+\frac{\partial}{\partial z_i}\Big(D^{\alpha0}_{1,i}(z)\,\hat L_\alpha\,\hat{\rho}_\mathrm{CQ}(z;t)\Big)\nonumber\\
&\;-i\big[\hat H(z),\hat{\rho}_\mathrm{CQ}(z;t)\big]+ D^{\alpha\beta}_0(z)\,\hat L_\alpha\,\hat{\rho}_\mathrm{CQ}(z;t)\,\hat L_\beta^\dagger-\frac{1}{2}D^{\alpha\beta}_0(z)\Big\{\hat L_\beta^\dagger \hat L_\alpha,\hat{\rho}_\mathrm{CQ}(z;t)\Big\}~.\label{eq:CQ_Pawula}
\end{align}
Here $z=(h,\pi)$ (and for field-theory applications the derivatives should be understood as functional derivatives).
The coefficients $D^{00}_{n,i_1\cdots i_n}(z)$, $D^{0\alpha}_{1,i}(z)$, $D^{\alpha0}_{1,i}(z)$, and $D^{\alpha\beta}_0(z)$ are the (generally $z$-dependent) moments appearing in the Kramers--Moyal expansion of the CQ channel.
Complete positivity further requires
\begin{align}
2D^{00}_2(z)\;\succeq\; D_1(z)\,D_0(z)^{-1}\,D_1(z)^\dagger,
\qquad
\big(\mathbb{I}-D_0(z)D_0(z)^{-1}\big)D_1(z)=0,
\label{eq:CQ_tradeoff_matrix}
\end{align}
where $D_0(z)^{-1}$ is the generalized inverse of the matrix $D_0^{\alpha\beta}(z)$,
$D_1$ is a matrix with entries $D^{0\alpha}_{1,i}(z)$ (and its adjoint contains $D^{\alpha0}_{1,i}(z)$),
and $D^{00}_2$ is a matrix in $(i,j)$ with entries $D^{00}_{2,ij}(z)$.
Furthermore, the zeroth moment $D_0^{\alpha\beta}(z)$ cannot vanish.
These constraints express the decoherence--diffusion trade-off required for a consistent (completely positive) hybrid evolution.

\bibliography{bib}

\end{document}